\documentclass[letterpaper,twocolumn,10pt]{article}

\usepackage[utf8]{inputenc}
\usepackage[english]{babel}
\usepackage[T1]{fontenc}
\usepackage{lmodern}
\usepackage[svgnames,x11names]{xcolor}
\usepackage{csquotes}

\PassOptionsToPackage{hidelinks,breaklinks=true}{hyperref}

\usepackage{usenix2019_v3}

\usepackage{multirow}

\usepackage{times}

\usepackage{caption}
\usepackage{paralist}

\usepackage[font={small}]{caption}
\usepackage[backend=biber,
bibstyle=numeric,
citestyle=numeric-comp,
sortcites=true,
maxbibnames=10,
]{biblatex}
\renewcommand*{\bibfont}{\footnotesize}

\usepackage[]{graphicx}

\usepackage[svgnames,x11names]{xcolor}

\usepackage{amsmath}
\usepackage{amssymb}
\usepackage{amsfonts}
\usepackage{amsfonts}
\usepackage{amsthm}
\usepackage{array}
\newcolumntype{$}{>{\global\let\currentrowstyle\relax}}
\newcolumntype{^}{>{\currentrowstyle}}

\usepackage{tabulary}
\usepackage{multirow}
\usepackage{pifont}
\PassOptionsToPackage{hyphens}{url}
\usepackage{float}
\usepackage{adjustbox}
\usepackage{booktabs}
\usepackage{wasysym}
\usepackage{xstring}
\usepackage{soul}
\usepackage[binary-units,group-separator={,}]{siunitx}
\usepackage{xspace}
\usepackage{cleveref}

\usepackage[disable]{todonotes}

\definecolor{lstgreen}{rgb}{0,0.7,0}
\usepackage{listings}
\lstdefinelanguage{solidity}{%
  morekeywords={new, true, false, function, return,
  switch, var, if, in, while, do, else, case, break, 
  contract, mapping, struct, library, 
  revert, assert, require, throw, sha3, keccak256,
  sstore, sload, delete},
  morekeywords=[2]{export, public, private, payable, view, returns},
  morekeywords=[3]{bool, uint, int, address, uint256, int256, int8, uint8,
    uint128, int128, bytes, bytes32, bytes4},
  morekeywords=[4]{msg, this, tx},
  sensitive=false,
  comment=[l]{//},
  morecomment=[s]{/*}{*/},
  morestring=[b]',
  morestring=[b]"
}
\lstdefinelanguage{evm}{%
  morekeywords={%
    push1, push2, push3, push32, not, and, or, jump, jumpi, swap1, swap2, push4, jumpdest, add, sub, mul, div, pop, invalid, dup1, dup2, dup3, dup8, stop, lt, iszero, revert, sha3, sload, sstore, 
  },
  sensitive=false,
  comment=[l]{//},
  morecomment=[s]{/*}{*/},
}

\definecolor{lstgreen}{rgb}{0,0.6,0}
\usepackage{listings}
\lstdefinestyle{plain}{%
  numbers=none,
  frame=none,
  xleftmargin=1pt,
  xrightmargin=1pt,
}

\usepackage{tikz}
\usetikzlibrary{shapes,arrows,positioning,shapes.multipart}

\newcommand{\circleone}{\ding{192}}
\newcommand{\circletwo}{\ding{193}}

\newcommand{\evm}{EVM\xspace}
\newcommand{\ether}{Ether\xspace}
\newcommand{\cfg}{control-flow graph\xspace}

\newcommand{\theDAO}{``TheDAO''\xspace}

\newcommand{\eInst}[1]{\texttt{#1}}

\newcommand{\funcname}[1]{\textit{#1}}

\newcommand{\hex}[1]{\(\mathtt{#1}\)}

\newcommand{\toolname}{\textsc{EVMPatch}\xspace}
\newcommand{\osiris}{Osiris\xspace}

\renewcommand{\paragraph}[1]{\noindent\textbf{#1}}

\makeatletter
\newcommand*{\rom}[1]{\expandafter\@slowromancap\romannumeral #1@}
\makeatother

\usepackage{balance}

\DeclareSIUnit\wei{wei}
\DeclareSIUnit\ether{eth}
\DeclareSIUnit\gas{gas}
\newcommand{\usd}[1]{\SI[round-precision=2,round-mode=places,round-integer-to-decimal]{#1}[US\$\ensuremath{\,}]{}}

\makeatletter%
\if@todonotes@disabled%
\else%
\fi
\makeatother

\crefname{section}{\S}{\S\S}
\crefname{subsection}{\S}{\S\S}
\crefname{subsubsection}{\S}{\S\S}

\DeclareUnicodeCharacter{0301}{*************************************}
\begin{document}
\date{}

\title{\Large \bf EVMPatch: Timely and Automated Patching of Ethereum Smart Contracts}

\newcommand{\authormark}[1]{$^{#1}$}

\author{%
  {\rm Michael Rodler}\\
  University of Duisburg-Essen
  \and
  {\rm Wenting Li}\\
  NEC Laboratories Europe
  \and
  {\rm Ghassan O. Karame}\\
  NEC Laboratories Europe
  \and
  {\rm Lucas Davi}\\
  University of Duisburg-Essen
}

\maketitle

\emph{A slightly shorter version of this paper will be published at USENIX Security Symposium 2021.}

\begin{abstract}

Recent attacks exploiting errors in smart contract code had devastating consequences thereby questioning the benefits of this technology.
It is currently highly challenging to fix errors and deploy a patched contract in time.
Instant patching is especially important since smart contracts are always online due to the distributed nature of blockchain systems.
They also manage considerable amounts of assets, which are at risk and often beyond recovery after an attack.
Existing solutions to upgrade smart contracts depend on manual and error-prone processes.
This paper presents a framework, called \toolname, to instantly and automatically patch faulty smart contracts.
\toolname\ features a bytecode rewriting engine for the popular Ethereum blockchain, and transparently/automatically rewrites common off-the-shelf contracts to upgradable contracts.
The proof-of-concept implementation of \toolname\ automatically hardens smart contracts that are vulnerable to integer over/underflows and access control errors, but can be easily extended to cover more bug classes.
Our extensive evaluation on \num{14000} real-world (vulnerable) contracts demonstrate that our approach successfully blocks attack transactions launched on these contracts, while keeping the intended functionality of the contract intact.
We perform a study with experienced software developers, showing that \toolname\ is practical, and reduces the time for converting a given Solidity smart contract to an upgradable contract by \SI{97.6}{\percent}, while ensuring functional equivalence to the original contract.

\end{abstract}

\section{Introduction}%
\label{sec:intro}

Smart contracts are used in modern blockchain systems to allow nearly arbitrary (Turing-complete) business logic to be implemented. They enable autonomous management of cryptocurrency or tokens and have the potential to revolutionize many business applications by removing the need for a trusted (potentially malicious) third party, e.g., in applications for payments, insurances, crowd funding, or supply chains. Due to their ease of use and the high monetary value (cryptocurrency) some of these contracts hold, smart contracts have become an appealing target for attacks. Programming errors in smart contract code can have devastating consequences as an attacker can exploit these bugs to steal cryptocurrency or tokens.

Recently, the blockchain community has witnessed several incidents due smart contract errors~\cite{dao-attack-analysis,parity-multisig-vuln}.
One especially infamous incident is the \theDAO\ reentrancy attack, which resulted in a loss of over 50 million US Dollars worth of Ether~\cite{daohackamount}.
This led to a highly debated hard-fork of the Ethereum blockchain.
Several proposals demonstrated how to defend against reentrancy vulnerabilities either by means of offline analysis at development time or by performing run-time validation~\cite{luu2016making,Tsankov2018-lq,sereum-ndss19,ecfchecker}.
Another infamous incident is the parity wallet attack~\cite{parity-multisig-vuln}.
In this case, an attacker moved a smart contract into a state, where the currency held by the contract could not be accessed anymore.
This resulted in a total of about \num{500000} Ether to be stuck in smart contracts due to an access control error~\cite{parity-postmorterm}.
Automatic detection of such access control vulnerabilities has been previously studied in the context of automated exploit generation~\cite{Krupp2018-teether,maian}.
Further, integer overflow bugs constitute a major vulnerability class in smart contracts.
Such bugs occur when the result of an arithmetic operation has a longer width than the integer type can hold~\cite{seicertint}.
According to a study by Torres et al.\ \cite{osiris} more than \num{42000} contracts suffer from an integer bug.
They especially affect so-called ERC-20 Token contracts, which are leveraged in Ethereum to create subcurrencies.
Interestingly, several of the disclosed vulnerabilities were actually exploited leading to substantial token and ether losses.

These attacks have fueled interest in the community to enhance the security of smart contracts.
In this respect, a number of solutions ranging from devising better development environments to using safer programming languages, formal verification, symbolic execution, and dynamic runtime analysis have been proposed in the last few years~\cite{luu2016making,zeus-ndss2018,sereum-ndss19}.
We point out that all these solutions only aim to prove the correctness or absence of a certain type of vulnerability~\cite{luu2016making,Tsankov2018-lq,zeus-ndss2018} and as such cannot be used to protect already deployed (legacy) contracts.
Although some contracts integrate upgrade mechanisms (see \Cref{sec:upgrade}), once a particular contract has been flagged as vulnerable, it is unclear how to automatically patch it and test the effectiveness of the patched contract.
Even though manually patching contracts on the source-code level seems plausible, the patch may unexpectedly break compatibility and make the upgraded contracts unusable.
For example, given the special storage layout design of Ethereum, the delegatecall-proxy pattern requires developers to ensure that the patched version of the contract is \emph{compatible} with the previously deployed version.
Even small changes like changing the ordering of variables in the source code can break this compatibility.
This additionally poses the challenge that developers must adhere to strict coding standards~\cite{zeppelinos-docs} and have to use the same exact compiler version.
As a result, patching smart contract errors is currently a \emph{time-consuming, cumbersome, and error-prone process}.
For instance, while patching the Parity multisig wallet contract, a vulnerability was introduced.
An attacker was able to become the owner of the newly deployed library contract.
This allowed the attacker to destroy the contract and break all contracts that depend on the multisig wallet library contract.
As a result, a considerable amount of Ether is now locked in those broken contracts~\cite{parity-postmorterm}.
On top of that, patching smart contract bugs is highly \emph{time-critical}. In contrast to errors discovered in PC or mobile software, smart contract errors are unique from an attacker's point of view as \emph{(1)~smart contracts are always online on the blockchain, (2)~they usually hold a significant amount of assets, and (3)~an attacker does not need to consider other environmental variables} (e.g., software and library version, network traffic analysis, spam or phishing mails to trigger the exploit through a user action).

\medskip%
\paragraph{Contributions.}
In this paper, we address the problem of automated and timely patching of smart contracts to aid developers to instantly take action on reported smart contract errors.
We introduce a novel patching framework (in \Cref{sec:architecture}) that features a \emph{bytecode-rewriter} for Ethereum smart contracts, is independent of the source programming language and works on unmodified contract code.
Our framework, dubbed \toolname, utilizes the bytecode-rewriting engine to ensure that patches are minimally intrusive and that the newly patched contract is compatible with the original contract.
In particular, our framework automatically \emph{replays transactions} on the patched contract to
\begin{compactenum}
	\item test the functional correctness of the patched contract with respect to previous transactions pertaining to the contract,
	\item identify potential attacks, i.e., developers can determine whether their vulnerable contract has been attacked in the past.
\end{compactenum}
\toolname\ uses a best effort approach to ensure the introduced patch does not break functionality by testing with previously issued transactions to the contract and optionally also developer provided unit tests.
While such a differential testing approach cannot provide a formal proof on the correctness of the patched contract, it works without requiring a formal specification.
Our experiments (see \Cref{sec:evaluation}) show that this approach is sufficient in practice to identify broken patches.

By applying patches on the bytecode level, \toolname\ is independent of the used programming language/compiler and compiler version.
That is, \toolname\ supports any off-the-shelf Ethereum smart contract code.
We employ bytecode writing to ensure minimally intrusive patches, that are compatible by design with the contract's storage layout, %
We argue that source-level patching is not easily usable in an automated patching process that we propose.
However, as for any approach working on either the binary or bytecode-level, we had to tackle several technical challenges (\Cref{sec:implementation}).
Furthermore, \toolname\ automatically converts the original contract to use the delegatecall-proxy pattern.
As such, \toolname\ is able to automatically deploy newly patched contracts in a fully automated way without requiring any developer intervention.

While in principle \toolname\ can support patching of different classes of vulnerabilities (see \Cref{sec:other-vulns}), our proof-of-concept implementation targets the two major classes of access control and integer overflow (\Cref{sec:intoverflow}) bugs.
The latter have been repeatedly exploited in high-value ERC-20 contracts~\cite{peckshieldadvisories}, whereas the former has been abused in the Parity wallet attack~\cite{parity-multisig-vuln}.

To evaluate \toolname\ in terms of performance, effectiveness, and functional correctness, we apply \toolname\ to \num{14000} real-world vulnerable contracts.
To this end, we used the patch testing component of the \toolname\ framework to re-play all existing transactions to the original contract on the patched contract.
This allows us to provide in-depth investigation of several actively exploited smart contracts, e.g., token burning and history of attack transactions (before and after public disclosure).
For a number of contracts we investigated in our evaluation, we found that \toolname\ would have blocked several attacks that happened after public disclosure of the vulnerability.
This shows that even though those contracts were officially deprecated, they were still used by legitimate users and exploited by malicious actors.
As such, there is an immediate need for tooling, as provided by \toolname, which allows the developers of smart contracts to efficiently patch their contracts.
Our evaluation also covers important practical aspects such as gas and performance overhead (i.e., the costs for executing transactions in Ethereum). The gas overhead for all our patched contracts was below \SI{0.01}{US\$} per transaction and the performance overhead negligible.

To assess the usefulness of \toolname, we conducted a sophisticated developer study\footnote{See \href{https://github.com/uni-due-syssec/evmpatch-developer-study}{github.com/uni-due-syssec/evmpatch-developer-study} for details} that focuses on comparing the usability of patching and deploying an upgradable contract with and without \toolname (\Cref{sec:evaldevstudy}).
Our study reveals that developers required \SI{62.5}{\minute} (median) to manually (without \toolname) convert a simple smart contract, which implements common Wallet functionality in about 80 lines of code, into an upgradable smart contract.
In spite of this considerable time, none of them performed a \emph{correct} conversion, leading to broken and potentially vulnerable contracts.
As such, this time measurements must be seen as a lower bound, as correctly converting a more complex contract will take even more time.
In contrast, the same task was performed by the developers using \toolname\ in \SI{1.5}{\minute} (median)---a reduction by \SI{97.6}{\percent}---while producing a correct upgradable contract.
\section{Background}%
\label{sec:background}

In this section, we provide background information on the Ethereum Virtual Machine (EVM), binary rewriting, and some common contract upgrade strategies.

\paragraph{EVM \& Smart Contracts: }
At the core of the Ethereum blockchain system lies a custom virtual machine, dubbed Etherum Virtual Machine (EVM), which executes Ethereum smart contracts.
EVM consists of a simple stack-based virtual machine with a custom instruction format.
Every instruction is represented as a one-byte opcode.
Arguments are passed on the data stack.
The only exception are the push instructions, which are used to push constants onto the stack.
These constants are encoded directly into the instruction bytes.
Furthermore, the \evm\ follows the Harvard architecture model and separates code and data into different address spaces.
In fact, the \evm\ features different address spaces for different purposes: %
the code address space, which contains a smart contract's code and is considered immutable, the storage address space for storing global state, and the memory address space for temporary data.

In the Ethereum network, a smart contract must be executed by every miner and every full node in the network to compute and verify the state before and after a block.
Ethereum features a mechanism to limit the execution time per smart contract and reward miners for executing smart contracts: the so-called \emph{gas}.
Every \evm\ instruction requires a certain gas budget to execute.
The transaction sender selects the price per gas unit in Ether and when a transaction is included into a block the corresponding Ether is transferred to the miner as a reward.
Minimizing the gas required for executing a contract is important as it indirectly minimizes the cost of operating a smart contract in Ethereum.

Smart contracts are developed in an object-oriented fashion, i.e., every smart contract has a defined interface of functions: the contract's ABI (Application Binary Interface).
Whenever a smart contract calls another smart contract, it utilizes one of the call instructions, such as \eInst{CALL} or \eInst{STATICCALL}.
The called contract will then process the provided input and update its own state accordingly.
A contract cannot directly access the state (i.e., the storage area) of other contracts and must always use function calls according to the ABI to retrieve any data from another contract.

In contrast to the regular \eInst{CALL} instruction, the \eInst{DELEGATECALL} instruction will execute the called contract's code in the context of the caller contract.
This instruction was introduced to implement \emph{library contracts}, i.e., common functionality can be deployed once to the blockchain and multiple contracts can rely on one library contract.
This means that the callee, i.e., the library contract, has full access to the state (the storage) and the Ether funds of the caller.
As such, a contract that utilizes a \eInst{DELEGATECALL} instruction must fully trust the callee.

\paragraph{Binary Rewriting: }
Binary rewriting is a well-known technique to instrument programs after compilation.
Binary rewriting has also been applied to retrofit security hardening techniques such as \emph{control-flow integrity}, to compiled binaries~\cite{Davi2012mocfi}, but also to dynamically apply security patches to running programs~\cite{Payer2013-pg}.
For binary rewriting on traditional architectures two flavors of approaches have been developed: static and dynamic rewriting.

Dynamic approaches~\cite{pin} rewrite code on-the-fly, i.e., while the code is executing.
This avoids imprecise static analysis on large binaries.
However, dynamic binary rewriting requires an intermediate layer, which analyzes and rewrites code at runtime.
Since the \evm\ does not support dynamic code generation or modification, it is not possible to apply this approach efficiently in Ethereum.
In contrast, static binary rewriting~\cite{dyninst,Laurenzano2010pebil} is applicable to Ethereum as it works completely offline.
It relies on static analysis to recover enough program information to accurately rewrite the code.

\paragraph{Contract Upgrade Strategies: }%
\label{sec:upgrade}
Ethereum treats the code of smart contracts as immutable once they are deployed on the blockchain\footnote{Except for the selfdestruction mechanism to kill a smart contract.}.
To remedy this, the community came up with strategies for deploying upgraded smart contracts~\cite{zeppelinos-blog-proxy-patterns,trailofbits-migration,upgradablecontracts}.
The most naive approach is to deploy the patched contract at a new address and migrate the state of the original contract to it.
However, state migration is specific to the contract and must be manually implemented by the developers of the contract.
It requires the contract developers to have access to all the internal state of the old contract, and
a procedure in the new contract to accept state transfers.
To avoid state migration, developers can also use a separate contract as a data storage contract, which is sometimes referred to as the eternal storage pattern~\cite{eternal-storage-eip,zeppelinos-blog-proxy-patterns}.
However, this adds additional gas overhead since every time the logic contract needs to access data it must perform a costly external call into the data storage contract.

A more common strategy is to write contracts with the proxy-pattern, with the most favorable version being the \textit{delegatecall-proxy} pattern.
Here, one smart contract is split into two different contracts, one for the code and one for data storage:
\begin{inparaenum}[i)]
	\item an immutable \emph{proxy contract}, which holds all funds and all internal state, but does not implement any business logic;
	\item a \emph{logic contract}, which is completely stateless and implements all of the actual business logic, i.e., this contract contains the actual code that governs the actions of the contract.
\end{inparaenum}
The proxy contract is the entry point of all user transactions.
It has immutable code and its address remains constant over the lifetime of the contract.
The logic contract implements the rules, which govern the behavior of the smart contract.
The proxy contract forwards all function calls to the registered logic contract using the \eInst{DELEGATECALL} instruction.
This instruction is used to give the logic contract access to all internal state and funds stored in the proxy contract.
To upgrade the contract, a new logic contract is deployed and its address is updated in the proxy contract.
The proxy contract then forwards all future transactions to the patched logic contract.
As a result, deploying upgraded contracts does not require any data migration, as all data is stored in the immutable proxy contract.
Moreover, the upgrading process is also \emph{transparent} to users, as the contract address remains the same.
Although existing blockchain platforms do not provide mechanisms to upgrade smart contracts, the usage of this proxy pattern allows \toolname\ to quickly upgrade a contract with negligible costs (in terms of gas consumption).

\section{Design of EVMPatch}%
\label{sec:architecture}

\begin{figure*}[t]
	\begin{center}
		\includegraphics[width=0.8\textwidth]{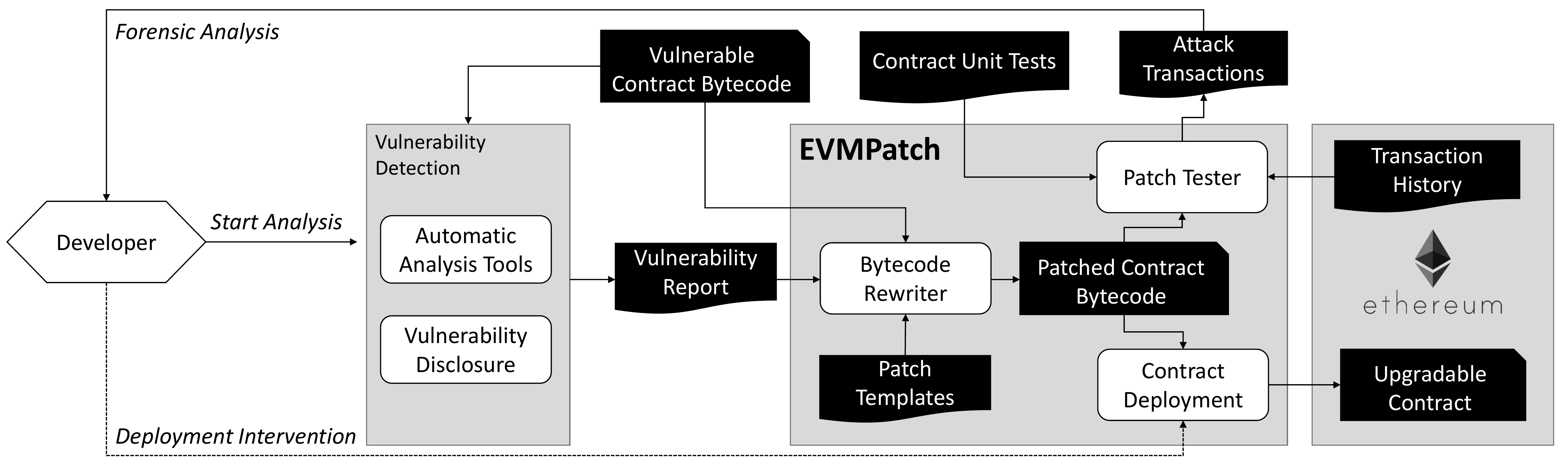}
	\end{center}
	\caption{Architecture of \toolname}%
	\label{fig:architecture}
\end{figure*}

In this section, we introduce the design of our automated patching framework to timely patch and harden smart contracts.
Our framework operates on unmodified smart contracts and is independent of the source code programming language, as it does not require source code.
At its core, our framework utilizes a bytecode rewriter to apply minimally intrusive patches to \evm\ smart contracts.
Combined with a proxy-based upgradable smart contract, this bytecode rewriting approach allows the developer to automatically introduce patches and deploy them on the blockchain.
One major advantage of this approach is that when new attack types are discovered or bug finding tools improve, the contract can be automatically re-checked, patched, and re-deployed in a short amount of time and with minimal developer intervention.
\toolname\ is typically executed on a developer's machine and is continuously running new and updated vulnerability detection tools.
This can also include dynamic analysis tools, which analyze transactions that are not yet included in a block, but already available to the Ethereum network.
Whenever one of the analysis tools discovers a new vulnerability, \toolname\ automatically patches the contract, tests the patched contract and deploys it.

\subsection{Design Choices}

The proxy-pattern makes it possible to easily deploy a patched smart contract in Ethereum.
However, it neither generates a patched version nor features functional tests on the patched contract.
\toolname\ fills this gap by providing a comprehensive framework and toolchain to
automatically and timely patch and test the effectiveness of the generated patch.

As shown in Table~\ref{tab:comp-rewriting}, there are two possible strategies for automatically generating a patch in Ethereum: static rewriting of source or \evm\ bytecode.
At first glance, \emph{source-code patching} seems to be the option of choice as developers have access to source code, they are able to inspect the source code changes, and can even do adjustments if the automated approach introduces undesired changes. However, in Ethereum, there is one major challenge when applying source code rewriting: one needs to carefully preserve the storage layout.
Otherwise, the patched contract will corrupt its memory and fail or (worse) introduce dangerous bugs.
Namely, some changes in the source code can break the contract compatibility, even though the changes do not break the logic of the contract.

\begin{table}[b]
	\caption{Comparison of rewriting strategies in Ethereum}%
	\label{tab:comp-rewriting}
	\renewcommand{\arraystretch}{1.2}
	\centering \small
	\begin{tabular}{ p{0.45\linewidth} | p{0.45\linewidth} }
		\toprule
		\textbf{Source Rewriting}                             & \textbf{Bytecode Rewriting}                      \\\midrule
		Corrupts storage-layout                               & Preserves storage layout                         \\
		Checking modifications by human analyst feasible      & Human analysis of bytecode changes challenging   \\
		Limited tool support for vulnerability analysis       & Easy integration of vulnerability analysis tools \\
		Patch testing based on prior transactions challenging & Easy patch testing with prior transactions       \\
		\bottomrule
	\end{tabular}
\end{table}

To put things into context, statically-sized variables are laid out contiguously in storage starting from address $0$; and contiguous variables with size less than \SI{32}{\byte} can be packed into a single \SI{32}{\byte} storage slot~\cite{storage-layout}.
As a result, any changes to re-order, add, or remove variables in the source-code may look harmless, but on the memory level, such changes will lead to mapping of variables to wrong and unexpected storage addresses.
In other words, changes in variable declaration corrupt the internal state of the contract, as the legacy contract and the patched contract have different storage layouts.

In contrast, \emph{bytecode rewriting} does not suffer from this deficiency as many bug classes only require changes on the level of \evm\ instructions (see \S\ref{sec:intoverflow}) avoiding any error-prone storage-layout changes. Another reason to opt for bytecode rewriting are existing smart contract vulnerability detection tools.
As of now, the majority of them operate on the EVM level~\cite{luu2016making,osiris,manticore-paper,Krupp2018-teether,sereum-ndss19} and report their findings on the EVM level.
A bytecode rewriting approach can exploit the reports of these analysis tools to directly generate an \evm\ bytecode-based patch. Finally, if source-code rewriting is utilized, the developer has limited possibilities to perform thorough testing on the effectiveness of the patched contract.
In particular, checking the patched contract against old transactions (including transactions that encapsulate attacks) are more feasible on bytecode level.
That is, transaction testing naturally would still require analysis on the bytecode level to reverse-engineer the attack transactions and how they fail against the patched contract.
Bytecode-rewriting allows developers to directly match the rewritten bytecode instructions to the attack transactions making forensic analysis feasible.
Given all these reasons, we decided to opt for bytecode rewriting.

\subsection{Framework Design}%
\label{sections:frameworkdesign}

Our framework depicted in Figure~\ref{fig:architecture} consists of the following major components: (1)~the \emph{vulnerability detection} engine consisting of automatic analysis tools and public vulnerability disclosures, (2)~\emph{bytecode rewriter} to apply the patch to the contract, (3)~the \emph{patch testing} mechanism to validate the patch on previous transactions, and (4)~the \textit{contract deployment} component to upload the patched version of the contract.
At first, the vulnerability detection engine identifies the location and type of the vulnerability.
This information is then passed to the bytecode rewriter, which patches the contract according to previously defined patch templates.
The patched contract is thereafter forwarded to the patch tester, which replays all past transactions to the contract.
That said, we do not only patch the contract, but we allow the developer to retrieve a list of transactions that exhibit a different behavior and outcome between the original and patched contract. These transactions serve as an indicator for potential attacks on the original contract. If the list is empty, our framework automatically deploys the patched contract instantly on the Ethereum blockchain. Next, we will provide a more detailed description of the four major components of our design.

\paragraph{Vulnerability Detection.}
Before being able to apply patches, our framework needs to identify and detect vulnerabilities. To do this, our framework leverages existing vulnerability detection tools such as~\cite{luu2016making,Tsankov2018-lq,Krupp2018-teether,maian,osiris,sereum-ndss19,ecfchecker}.
For vulnerabilities that are not detected by any existing tool, we require that a developer or a security consultant creates a vulnerability report.
In our system, the vulnerability detection component is responsible to identify the exact address of the instruction, where the vulnerability is located, and the type of vulnerability.
This information is then passed to the bytecode rewriter, which patches the contract accordingly.

\paragraph{Bytecode Rewriter.}
In general, static binary rewriting techniques are well suited for applying patches in Ethereum since smart contracts have comparably small code size: typically in the range of about 10 KiB.
Furthermore, \evm\ smart contracts are always statically linked to all library code.
It is not possible for a contract to dynamically introduce new code into the code address space.
This makes the reliance on binary rewriting techniques simpler compared to traditional architectures, where dynamically linked libraries are loaded at runtime.
However, some smart contracts still utilize a concept similar to dynamically linked libraries: dedicated EVM call instructions allow a contract to switch to a different code address space.
We tackle this peculiarity by applying our bytecode rewriter to both the contract itself and the library contract.

The stack-based architecture of the \evm\ requires special attention when implementing a patch:
all address-based references to any code or data in the code address space of the smart contract must be either preserved or updated when new code is inserted into the code address space.
Such references cannot be easily recovered from the bytecode. To tackle this challenge, \toolname\ utilizes a trampoline-based approach for adding new \evm\ instructions into empty code areas. The implementation details will be described in \Cref{sec:implementation}.

To implement a patch, the bytecode rewriter processes the bytecode of the vulnerable contract as well as the vulnerability report. The rewriting is based on a so-called patch template which is selected according to the vulnerability type and adjusted to work with the given contract.

\paragraph{Patch Templates.}
In \toolname, we utilize a template-based patching approach: for every supported class of vulnerabilities, a patch-template is integrated into \toolname.
This patch template is automatically adapted to the contract that is being patched.
We create generic patch templates such that they can be easily applied to all contracts.
\toolname\ \emph{automatically} adapts the patch template to the contract at hand by replacing contract-specific constants (i.e., code addresses, function identifier, storage addresses).
Patch templates for common vulnerabilities, such as integer overflows, are shipped as part of \toolname, and a typical user of \toolname\ will never interact with the patch templates.
However, optionally, a smart contract developer can also inspect or adapt existing patch templates or even create additional patch templates for vulnerabilities that are not yet supported by \toolname.

\paragraph{Patch Tester.}
As smart contracts directly handle assets (such as Ether), it is critical that any patching process does not impede the actual functionality of a contract.
As such, any patch must be tested thoroughly.
To address this issue, we introduce a patch testing mechanism which is based (1)~on the transaction history recorded on the blockchain and (2)~optional developer supplied unit tests.
At this point, we exploit the fact that any blockchain system records all previous executions of a smart contract, i.e., transactions in Ethereum.
In our case, the patch tester re-executes all existing transactions and optionally any available unit test and verifies that all transactions of the old legacy and the newly patched contract behave consistently.
The patch tester detects any behavioral discrepancy between the old legacy and the newly patched contract and reports a list of transactions with differing behavior to the developer. That said, as a by-product, our patch testing mechanism can be used as a forensic attack detection tool.
Namely, while executing the patching process, the developer will also be notified of any prior attacks that abuse any of the patched vulnerabilities and can then act accordingly.
In case both versions of the contract behave the same way, the patched contract can be automatically deployed.
Otherwise, the developer must investigate the list of suspicious transactions and thereafter invoke the contract deployment component to upload the patched contract.
The list of suspicious transactions may not only serve as an indicator of potential attacks, but may reveal that the patched contract is not functionally correct, i.e., the patched contract shows a different behavior on benign transaction.
In \Cref{sec:intoverflow}, we provide a thorough investigation on real-world, vulnerable contracts to demonstrate that \toolname\ successfully applies patches without breaking the original functionality of the contract.

\paragraph{Contract Deployment.}
As discussed in \Cref{sec:upgrade}, the delegatecall-proxy based upgrade scheme is the option of choice to enable instant contract patching.
Thus, \toolname\ integrates this deployment approach utilizing a proxy contract as the primary entry point for all transactions with a constant address.
Before the first deployment, \toolname\ transforms the original unmodified contract code to utilize the delegatecall-proxy pattern.
This is done by deploying a proxy contract, which is immutable and assumed to be implemented correctly\footnote{\toolname\ comes with a well audited default proxy contract that is only 80 lines of Solidity code.}.
The original bytecode is then converted to a logic contract using the bytecode rewriter with only minor changes to the original code.
The logic contract is then deployed alongside the proxy contract.

\paragraph{Patch Deployment.}
Finally, when the contract is patched and after the patch is tested by the patch tester component, \toolname\ can deploy the newly patched contract.
Our upgrade scheme deploys the newly patched contract code to a new address and issues a dedicated transaction to the previously deployed proxy contract, which switches the address of the logic contract from the old vulnerable version to the newly patched version.
Any further transactions are now handled by the patched logic contract.

\paragraph{Human Intervention.}
\toolname\ is designed to be fully automated.
However, there are a few scenarios, where developer intervention is needed if
\begin{inparaenum}[(1)]
  \item the vulnerability report relates to a bug class that is not yet supported by \toolname, or
  \item the patch tester reports at least one transaction that fails due to the newly introduced patch and the failing transaction is not a known attack transaction,
  \item the patch tester reports that at least one known attack transaction is not prevented by the newly introduced patch.
\end{inparaenum}

If a bug class is not supported, \toolname\ informs the developer about the unsupported vulnerability class.
Since \toolname\ is extensible, it easily allows developers to provide custom patch templates thereby allowing quick adaption to new attacks against smart contracts.
More specifically, \toolname\ supports multiple formats for custom patch templates: EVM instructions, a simple domain-specific language that resembles Solidity expressions and allows developers to enforce pre-conditions on functions (similar to Solidity modifiers).
We performed a developer study in Section~\ref{sec:evaldevstudy} to demonstrate that writing a patch template is feasible and more successful than manually patching a contract.

If the patch tester finds a new failing transaction, the developer has to analyze whether a new attack transaction has been discovered or a legitimate transaction has failed.
For a newly discovered attack transaction, \toolname\ adds this transaction to the list of attacks and proceeds.
Otherwise, the developer investigates why the legitimate transaction failed.
As our evaluation in \Cref{sec:evalfpfn} shows, such cases typically occur due to inaccurate vulnerability reports, i.e., wrongly reported vulnerabilities rather than faulty patching.
Thus, the developer can simply blacklist the wrongly reported vulnerable code locations to avoid patching at these locations.

These manual interventions typically only need quick code reviews or debugger sessions.
We believe even moderately experienced Solidity developers can perform these tasks as no detailed knowledge about the underlying bytecode rewriting system is needed (see also \Cref{sec:evaldevstudy} on our developer study).
As such, \toolname\ positions itself as a tool to enable more developers to securely program and operate Ethereum smart contracts.

\section{EVMPatch Implementation}%
\label{sec:implementation}

In this section, we describe the implementation of \toolname: in \Cref{sec:challenges}, we discuss engineering challenges for bytecode rewriting in Ethereum. Thereafter, we desribe the implementation of the bytecode rewriter (\Cref{sec:rewriter-impl}), the patch testing feature (\Cref{sec:patch-testing-impl}), and the contract deployment mechanism (\Cref{sec:patch-deployment-impl}). We conclude this section with a discussion on possible applications regarding smart contract errors in \Cref{sec:other-vulns}.

\subsection{Challenges of Bytecode Rewriting}%
\label{sec:challenges}

There are several unique challenges that must be solved when rewriting \evm\ bytecode: we need to handle static analysis of the original \evm\ bytecode, and tackle several particularities of Solidity contracts and the \evm.

Similar to traditional computer architectures, \evm\ bytecode uses addresses to reference code and data constants in the code address space.
Hence, when modifying the bytecode, the rewriter must ensure that address-based references are correctly adjusted.
To do so, a rewriter typically employ two static analysis techniques: control-flow graph (CFG) recovery and subsequent data-flow analysis.
The latter is necessary to determine which instructions are the sources of any address constants utilized in the code.
For the \evm\ bytecode, two classes of instructions are relevant in this context: code jumps and constant data references.

\paragraph{Code Jumps.}
The \evm\ features two branch instructions: \eInst{JUMP} and \eInst{JUMPI}.
Both take the destination address from the stack.
Note that function calls inside the same contract also leverage \eInst{JUMP} and \eInst{JUMPI}. That said, there is \emph{no explicit difference} between local jumps inside a function and calls to other functions. The EVM also features dedicated call instructions, but these are only used to transfer control to a completely separate contract. Hence, they do not require modification when rewriting the bytecode.%

\paragraph{Constant Data References.}
The so-called \eInst{CODECOPY} instruction is leveraged to copy data from the code address space into the memory address space.
A common example use-case are large data constants such as strings.
Similar to the jump instructions, the address from which memory is loaded is passed to the \eInst{CODECOPY} instruction via the stack.%

Handling both types of instructions is challenging due to the stack-based architecture of the \evm.
For instance, the target addresses of jump instructions are always provided on the stack.
That is, every branch is indirect, i.e., the target address cannot be simply looked up by inspecting the jump instruction. Instead, to resolve these indirect jumps, one needs to deploy data-flow analysis techniques to determine where and which target address is pushed on the stack.
For the majority of these jumps, one can analyze the surrounding basic block\footnote{A basic block is sequence of EVM instructions that terminate in a branch. The branch connects one basic block to subsequent basic blocks in the CFG of the \evm\ code.} to trace back where the jump target is pushed on the stack.
For example, when observing the instructions \texttt{PUSH2 0xdb1; JUMP}, we can recover the jump target by retrieving the address (\hex{0xdb1}) from the push instruction.

However, many contracts contain more complicated code patterns, primarily because the Solidity compiler also supports calling functions internally without utilizing a call instruction.
Recall that, in the \evm, a call instructions perform similarly to remote-procedure calls.
To optimize code size and facilitate code re-use, the Solidity compiler introduced a concept where functions are marked as \emph{internal}.
These functions cannot be called by other contracts (private to the contract) and follow a different calling convention.
Since there are no dedicated return and call instruction for internal functions, Solidity utilizes the jump instruction to emulate both.
As such, a function return and a normal jump cannot be easily distinguished.
This makes it challenging to (1)~identify internal functions and (2)~build an accurate \cfg\ of the contract.

When rewriting an \evm\ smart contract, both the jump instructions and the codecopy instruction need to be considered in the bytecode rewriter.
The obvious strategy to rewrite smart contracts is to fix-up all constant addresses in the code to reflect the new addresses after inserting new instructions or removing old instructions.
However, this strategy is challenging because it requires accurate \cfg\ recovery and data-flow analysis, which needs to deal with particularities of \evm\ code, such as internal function calls.
In the research area of binary rewriting of traditional architectures, a more pragmatic approach has been developed: the so-called trampoline concept~\cite{Laurenzano2010pebil,Davi2012mocfi}.
We utilize this approach in our rewriter and avoid adjusting addresses. Whenever our rewriter must perform changes to a basic block, e.g., inserting instructions, our rewriter replaces the basic block with a trampoline that immediately jumps to the patched copy.
Hence, any jump target in the original code stays the same and all data constants are kept at their original addresses.
We describe this process in more detail in the subsequent section.

\subsection{Bytecode Rewriter Implementation}%
\label{sec:rewriter-impl}

We implemented a trampoline-based rewriter in Python and utilize the \emph{pyevmasm}\footnote{\href{https://github.com/crytic/pyevmasm}{github.com/crytic/pyevmasm}} library for disassembling and assembling raw \evm\ opcodes.
Our trampoline-based bytecode rewriter works on the basic block level.
When an instruction needs to be instrumented, the whole basic block is copied to the end of the contract.
The patch is then applied to this new copy.
The original basic block is replaced with a trampoline, i.e., a short instruction sequence that immediately jumps to the copied basic block.
Whenever the contract jumps to the basic block at its original address, the trampoline is invoked redirecting execution to the patched basic block by means of a jump instruction.
To resume execution, the final instruction of the instrumented basic block issues a jump back into the original contract code.
While the trampoline-based approach avoids fixing up any references, it introduces additional jump instructions.
However, as we will show, the gas cost associated with these additional jumps is negligible in practice (see \Cref{sec:intoverflow}).

To ensure correct execution, we must still compute at least a partial \cfg, starting from the patched basic block.
This is necessary to recover the boundaries of the basic blocks that are patched and the following basic blocks that are connected by a so called fall-through edge.
Not all basic blocks terminate with an explicit control-flow instruction: %
Whenever a basic block ends with a conditional jump instruction (\eInst{JUMPI}) or simply does not end with a control-flow instruction, there is an implicit edge (i.e., fall-through) in the control-flow graph to the instruction at the following address.

\paragraph{Handling Fall-Through Edge.}
To handle the fall-through edge, two cases must be considered.
When the basic block targeted by the fall-through edge starts with a \eInst{JUMPDEST} instruction, the basic block is marked as a legitimate target for regular jumps in the \evm.
In this case, we can append an explicit jump to the rewritten basic block at the end of the contract and ensure that execution continues at the beginning of the following basic block in the original contract code.
In case that the following basic block does not begin with a \eInst{JUMPDEST} instruction, the \evm\ forbids explicit jumps to this address.
In the \cfg, this means that this basic block can only be reached with a fall-through edge.
To handle this case, our rewriter copies the basic block to the end of the contract right behind the rewritten basic block constructing another fall-through edge in the \cfg\ of the rewritten code.

\begin{figure}[t]
	\begin{center}
		\includegraphics[width=\linewidth]{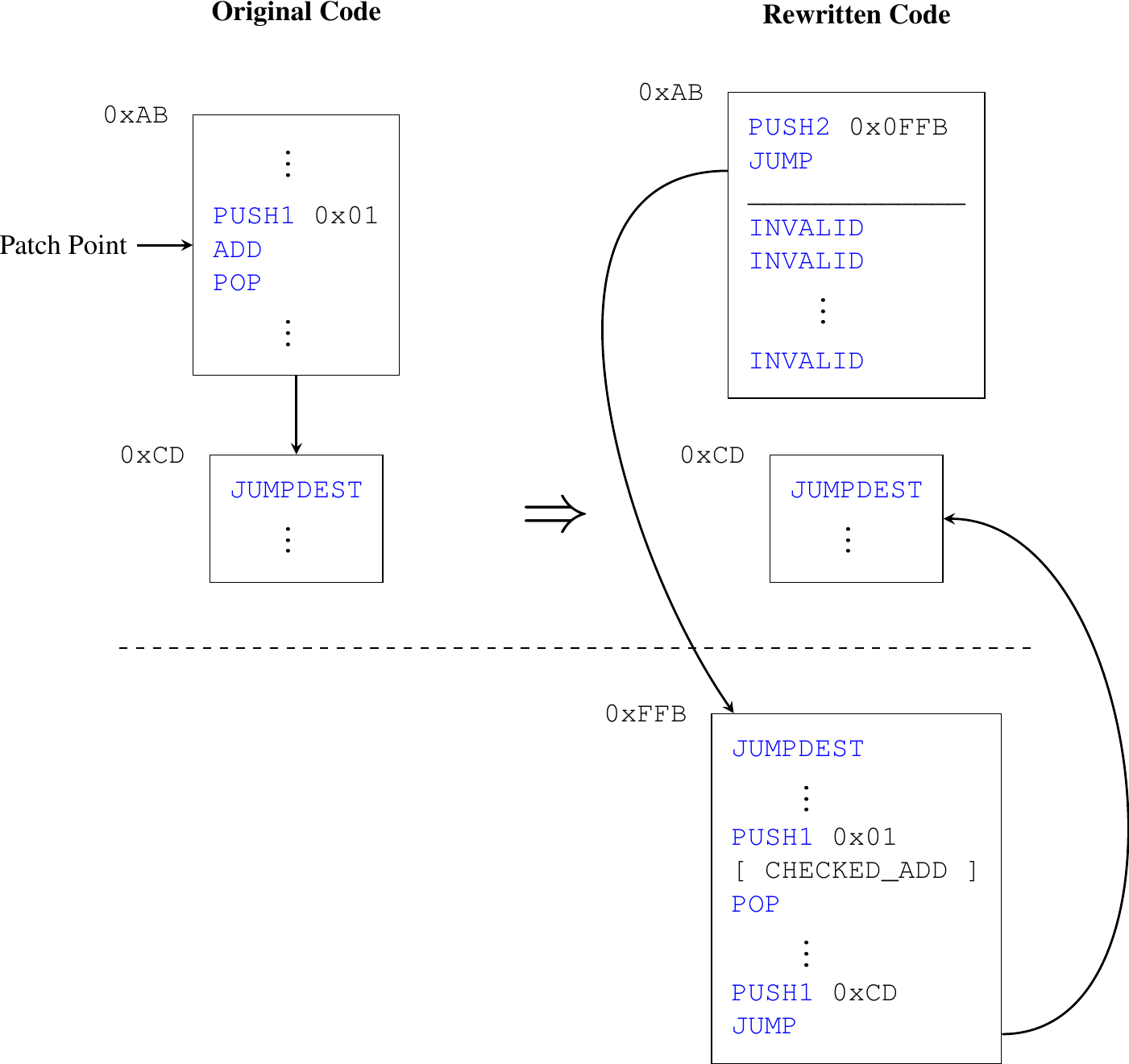}
	\end{center}
	\caption{Control-flow graph of original and rewritten code.
	}%
	\label{fig:rewrittencode}
\end{figure}

Figure~\ref{fig:rewrittencode} shows an example for how our rewriter changes the \cfg\ of the original contract.
The \eInst{ADD} instruction is replaced with a checked add routine that additionally performs integer overflow checks.
We call the address of the \eInst{ADD} instruction the \emph{patch point}.
The basic block, which contains the patch point, is replaced with a trampoline.
In this case, it immediately jumps to the basic block at \hex{0xFFB}.
This basic block, which is placed at the end of the original contract, is a copy of the original basic block at \hex{0xAB}, but with the patch applied.
Since the basic block is now at the end of the contract, the bytecode rewriter can insert, change, and remove instructions in the basic block without changing any address in code that is located at higher-numbered addresses.
We fill the rest of the original basic block with the \eInst{INVALID} instruction to ensure the basic block has the exact same size as the original basic block.
The basic block at \hex{0xCD} is connected to the prior basic block by means of a fall-through edge.
However, this basic block starts with a \eInst{JUMPDEST} instruction and as such is a legitimate jump target.
Hence, the rewriter then appends a jump to the patched basic block at \hex{0xFFB} which ensures execution continues in the original contract's code at address \hex{0xCD}.

\paragraph{Adapting to EVM.}
The \evm\ has some particularities that must be considered when implementing a bytecode rewriter.
Namely, the \evm\ enforces some separation of code and data in the code address space.
\evm\ implementations prevent jumps into the data constants that are embedded into \eInst{PUSH} instructions.
The constant operands of the push instructions follow directly after the byte of the push instruction opcode.
Such a constant operand can accidentally include the byte for the \eInst{JUMPDEST} instruction.
Then, the constant would be a legitimate jump target and a new unintended instruction sequence would occur.
To avoid such unintended instruction sequences, \evm\ implementations perform a linear sweep over the code section to find all push instructions.
The constants that are part of those push instructions are then marked as data and therefore as invalid jump targets, even if they contain a byte equivalent to the \eInst{JUMPDEST} instruction.
However, due to performance reasons, \evm\ implementations ignore control-flow information when marking data.
As such, the push instructions opcode byte itself can be part of some data constant, such as a string or other binary data.
For this reason, smart contract compilers accumulate all data constants at addresses strictly larger than any reachable code, avoiding any conflicts between the generated code and data encoded into the code address space.
However, our trampoline-based rewriter does append code behind the data constants of the smart contracts.
To avoid that code appended by the rewriter is accidentally marked as an invalid jump destination due to a preceding push opcode byte, we carefully insert padding between the data of the original contract and the newly appended code.

\paragraph{Applicability of Trampoline Approach.}
The trampoline-based approach to rewriting requires only minimal code analysis and works for most use cases.
However, this approach faces two problems.
First, instructions can only be patched in basic blocks that are large enough (in terms of size in bytes) to also contain the trampoline code.
However, a typical trampoline requires 4 to 5 bytes and typically basic blocks that perform some meaningful computation are large enough to contain the trampoline code.
Second, due to the copying of basic blocks the code size increases depending on the basic block that is patched thereby increasing deployment cost.
However, our experiments show that the overhead during deployment is negligible (on average \usd{0.02} per deployment, see \Cref{sec:intoverflow}).

\paragraph{No reliance on accurate \cfg.}
Recovering an accurate \cfg\ given only EVM bytecode is a challenging and open problem. %
However, our trampoline based approach does not require an accurate and complete \cfg.
Instead, we only need to recover basic block boundaries given the program counter of the instruction, where the patch needs to be applied.
In doing so, recovering the basic block boundaries is tractable, since the \evm\ has an explicit marker for basic block entries (i.e., the \texttt{JUMPDEST} pseudo-instruction).
Furthermore, our rewriter only needs to recover the end of the basic block and any following basic blocks that are connected via fallthrough edges in the \cfg.

\subsection{Patch Testing}%
\label{sec:patch-testing-impl}

While the insertion of trampolines into the original code does not change the functionality of the contract, the patch template itself can perform arbitrary computations and could potentially violate the semantics of the patched contract.
To test the patched contract, \toolname\ utilizes a differential testing approach. 
That is, we re-execute all transactions of the contract to determine if the behavior of the original, vulnerable code and the newly, patched code differ.
\toolname\ utilizes past transactions to the contract retrieved directly from the blockchain.
If the contract comes with unit tests, \toolname\ also utilizes the unit tests to test the newly patched contract.
This differential testing approach cannot guarantee formal correctness of the contract.
Contracts with a low number of available transactions are prone to low test coverage.
However, our experiments (see \Cref{sec:evaluation}) show that the differential testing approach works well enough in practice to show that the patches do not break functionality.
Given the availability of a formal specification of the contract's functionality, \toolname\ could also leverage a model checker to validate a patched contract more rigorously.

During differential testing, we first retrieve a list of transactions to the vulnerable contract from the blockchain.
Second, we re-execute all those transactions and retrieve the execution trace for each transaction.
Then, we then re-execute the same transactions, but replace the code of the vulnerable contract with the patched contract code, to obtain the second execution trace.
We use a modified Ethereum client, based on the popular \emph{go-ethereum} client\footnote{We utilized version 1.8.27-stable-3e76a291}, since the original client does not support this functionality.
Finally, we compare both execution traces and the patch tester produces a list of transactions, where the behavior differs.
If there are no such transactions, then we assume that the patch does not inhibit the functionality of the contract and proceed with deploying the patched contract.

The execution traces of the original and patched contracts are never equal since patching changes control flow and inserts instructions.
Hence, we examine only potentially state-changing instructions, i.e., instructions that either write to the storage area (i.e., a \eInst{SSTORE}) or transfer execution flow to another contract (e.g., a \eInst{CALL} instruction).
We then compare the order, parameters, and result of all state-changing instructions and find the first instruction where the two execution traces differ.
Currently, we assume that the introduced patches do not result in any new state-changing instructions.
This assumption holds for patches that introduce input-validation code and revert when invalid input is passed.
However, the trace difference computation can be adapted to become aware of potential state changes that a patch introduces.%
Reported transactions that fail in the code, which is part of the patch, are marked as potential attack transactions.
If the reported transaction failed due to out-of-gas in the patched code, we re-run the same transaction with an increased gas budget.
We issue a warning since users will have to account for additional gas cost introduced by the patch.
Finally, the developer must examine the reported transactions to decide whether the given list of transactions are legitimate or malicious.
As a side-effect, this makes our patch tester an attack detection tool for the vulnerable contract allowing developers to quickly find prior attack transactions.

\subsection{Deployment of Patched Contracts}%
\label{sec:patch-deployment-impl}

As described in \Cref{sec:architecture}, \toolname\ utilizes the delegatecall-proxy based upgrade pattern to deploy the patched contract.
To achieve this, \toolname\ splits the smart contract to two contracts: a proxy contract and a logic contract.
The proxy contract is the primary entry point and stores all data.
By default, \toolname\ utilizes a proxy contract that is shipped with \toolname.
However, \toolname\ can also re-use existing upgradable contracts, such as contracts developed with the ZeppelinOS framework~\cite{zeppelinos-docs}.
Users interact with the proxy contract, which is located at a fixed address.
To facilitate the upgrade process, the proxy contract also implements functionality to update the address of the logic contract.
To prevent malicious upgrades, the proxy contract also stores the address of an owner, who is allowed to issue upgrades.
The upgrade then simply consists of sending one transaction to the proxy contract, which will (1)~check whether the caller is the owner and (2)~update the address of the logic contract.

The proxy contract retrieves the address of the new logic contract from storage and simply forwards all calls to that contract.
Internally, the proxy contract utilizes the \eInst{DELEGATECALL} instruction to call into the logic contract.
This allows the logic contract to gain full access to the storage memory area of the proxy contract thereby allowing access to the persistent data without any additional overhead.

\subsection{Possible Applications}%
\label{sec:other-vulns}

The bytecode rewriter takes a patch template, which is specified as short snippet of \evm\ assembly language.
This template is then specialized according to the patched contract and relocated to the end of the patched contract.
This template-based approach to patch generation allows to specify multiple generic patches to address whole classes of vulnerabilities.
In the following, we list possible vulnerability classes that can immediately benefit from our framework.

\paragraph{Improper access control}
to critical functions can be patched by just inserting a check at the beginning of a function to verify that the caller is a certain fixed address or equal to some address stored in the contract's state.
Detection tools to handle this vulnerability have been investigated in prior work~\cite{Krupp2018-teether,maian}.

\paragraph{Mishandled exceptions}
can occur when the contract uses a low-level call instruction, where the return value is not handled automatically, and the contract does not properly check the return value~\cite{luu2016making}.
This issue can be patched by inserting a generic return-value check after such a call instructions.

\paragraph{Integer bugs} are highly likely to occur when dealing with integer arithmetic since Solidity does not utilize checked arithmetic by default.
This has resulted in many potentially vulnerable contracts being deployed and some being actively attacked~\cite{osiris,peckshieldadvisories}.
Given the prevalence of these vulnerabilities, we discuss in the next section how to automatically patch integer overflow bugs using \toolname.

In what follows, we demonstrate the effectiveness of \toolname\ by applying it to the two major bug classes of access control errors and integer bugs.

\section{Evaluation of \toolname}%
\label{sec:intoverflow}

In this section, we report the evaluation results of \toolname\ in patching two prominent types of bugs: (1)~access control bugs, and (2)~integer bugs (over-/underflow).

\subsection{Patching Access Control Bugs}

The Parity MultiSig Wallet is a prominent example for access control errors~\cite{parity-attack,parity-multisig-vuln}. This contract implements a wallet that is owned by multiple accounts.
Any action taken by the wallet contract must be authorized by at least one of the owners.
However, the contract suffered from a fatal bug that allowed anyone to become the sole owner because the corresponding functions \funcname{initWallet}, \funcname{initMultiowned}, and \funcname{initDayLimit} did not perform any access control checks.

Figure~\ref{fig:parity} shows the patched source code which adds the \textit{internal} modifier to the functions \funcname{initMultiowned} and \funcname{initDayLimit} (marked with \circleone\ in Figure~\ref{fig:parity}).
This modifier makes these two functions inaccessible via the outside interface of the deployed contract.
Furthermore, the patch adds the custom modifier \funcname{only\_uninitialized}, which checks whether the contract was previously initialized (marked with \circletwo).

\begin{figure}[t]
	\begin{center}
		\input{figures/parity_source.tex}
	\end{center}
	\caption{Source code of patched Parity Multisig Wallet.}%
	\label{fig:parity}
\end{figure}

The developers originally introduced a new vulnerability while deploying the patched the contract, which was actively exploited~\cite{parity-postmorterm}.
In contrast, because \toolname\ performs bytecode rewriting, it would have immediately generated a securely patched version of the contract and would have deployed it automatically in a secure manner.

Consider Figure~\ref{fig:parity-patch} which shows a customized patch in the domain-specific language employed by \toolname\ to specify patches.
As such, we insert a patch at the beginning of the \funcname{initWallet} function that checks whether the condition \verb+sload(m_numOwners) == 0+ holds, i.e., whether the contract is not yet initialized.
If this does not hold, the contract execution will abort with a \eInst{REVERT} instruction.
Note that here an explicit \texttt{sload} needs to be used to load variables from storage and the expression is logically inverted from the patch in \Cref{fig:parity}, since this patch essentially inserts a Solidity \texttt{require} statement.
Furthermore, two other publicly accessible functions need to be removed from the public function dispatcher.
The patch shown in \Cref{fig:parity-patch} combines two existing patch templates provided by \toolname.
First, the \emph{add require} patch template enforces a pre-condition before a function is entered.
Second, the \emph{delete public function} patch template removes a public function from the dispatcher, effectively marking the function as internal.

\begin{figure}[t]
  \begin{center}
    \input{figures/parity_patch_script.tex}
  \end{center}
  \caption{Customized Patch for Partity Multsig Wallet.}%
  \label{fig:parity-patch}
\end{figure}

\paragraph{Evaluation Results.}%
We verified that the patched contract is no longer exploitable by deploying a patched version of the \funcname{WalletLibrary} contract against the attack.
Further, we compare a source-level patch with the patch applied by \toolname.
\Cref{tab:paritycomparison} shows an overview of the results.
\toolname only increases contract size by \SI{25}{\byte}. The additional gas cost of the \funcname{initWallet} function is only \SI{235}{\gas}, i.e., \num{0.00006}~USD per transaction for \(235.091\) USD/ETH and a typical gas price of \SI{1}{\giga\wei}.
This demonstrates that \toolname can efficiently and effectively insert patches for access control bugs.%

\begin{table}[b]
\begin{center}
	\caption{Overhead of access control patch.}%
  \label{tab:paritycomparison}
	\small
	\begin{tabular}{c c c c}
		\toprule
		Version        & Bytes & Size Increase  & Gas Increase  \\
		\midrule
		Original       & \num{8290}   & \num{0} \%     & 0            \\
		Source-Patched & \num{8201}   & \num{-1.07} \% & 226          \\
		\toolname'ed   & \num{8315}   & \num{0.3} \%   & 235          \\
		\bottomrule
	\end{tabular}
\end{center}
\end{table}

\subsection{Patching Integer Bugs}%

Typical integer types are bound to a minimum and/or maximum size due to the fixed bit-width of the integer type.
However, programmers often do not pay sufficient attention to the size limitation of the actual integer type potentially causing integer bugs. %
Fortunately, several high-level programming languages (Python, Scheme) are able to avoid integer bugs since they leverage arbitrary precision integers with virtually unlimited size.
However, the de-facto standard programming language for smart contracts, namely Solidity, does not embed such a mechanism. %
This leaves the burden of handling integer overflows completely on the developer who needs to either manually implement overflow checks or properly utilize the SafeMath library to safely perform numeric operations~\cite{safemath}. While common, the former is obviously error-prone.
For instance, multiple vulnerabilities in ERC-20 token contracts were recently unveiled~\cite{burnOverflowHexagon,batchoverflow,SCAmultioverflow}. These contracts manage subcurrencies, so-called tokens, on the Ethereum blockchain.
Such tokens can deal with large amounts of currency since they track the token balance of every token owner and mediate the exchange of tokens and Ether.
Figure~\ref{fig:batchoverflow} shows an excerpt of the \emph{BEC} token contract's code that exemplifies such integer overflow vulnerabilities.
When computing the total amount in Line~6, an unchecked integer multiplication is used allowing an attacker to provide a very large \emph{\_value}.
As a consequence, the \emph{amount} variable will be set to a small amount.
This effectively bypasses the balance check in Line~11
allowing the attacker to transfer a large amount of tokens to an attacker-controlled account.
Recently, similar vulnerabilities have been discovered in over \num{42000} contracts~\cite{osiris}.

\begin{figure}[t]
	\begin{center}
		\input{figures/batchoverflow_sol}
	\end{center}
	\caption{Integer overflow bug reported by PeckShield~\cite{batchoverflow}.}%
	\label{fig:batchoverflow}
\end{figure}

We developed patch templates for detecting integer overflows and underflows for the standard \evm integer width, i.e., unsigned \SI{256}{\bit} integers.
For integer addition, subtraction, and multiplication, these templates add checks inspired by secure coding rules in the C programming language~\cite{seicertint} and the \emph{SafeMath}~\cite{safemath} Solidity library.
When a violation is detected, \toolname\ issues an exception to abort and roll back the current call to the contract.

\subsubsection{Evaluation Results}%
\label{sec:evaluation}

To verify the correctness of the patches generated by our bytecode rewriter, we utilized the state-of-the-art integer detection tool \osiris~\cite{osiris} for vulnerability detection. After analyzing \num{50535} unique contracts in the first \num{5000000} blocks of the Ethereum blockchain, \osiris\ detects at least one integer overflow vulnerability in \num{14107} contracts. Using \toolname, we were able to successfully patch almost all of these contracts automatically. More specifically, we could not patch 33 contracts amongst the 14107 investigated contracts because the basic block, where the detected vulnerability was located is too small for the trampoline code.

\begin{table*}[h]
	\begin{center}\small
		{\setlength{\tabcolsep}{4pt}\small
			\begin{tabular}{*{11}{c}}
				\toprule
				\multirow{2}{*}{Contract} & \multirow{2}{*}{CVE} & \multirow{2}{*}{\# Patches} & \multicolumn{2}{c}{\# Transactions} & \multicolumn{2}{c}{Overhead (gas)} & \multicolumn{2}{c}{Code Size Increase (B)} & \multicolumn{2}{c}{Additional Cost RW (US\$)}                                                         \\\cline{4-11}
				                          &                      &                             & Total                               & Attacks                            & RW                                         & SM                                         & RW             & SM           & per TX   & per Upgrade \\
				\midrule
				BEC~\cite{bec-token}      & 2018-10299           & \num{1}                     & \num{424229}                        & 1                                  & 83                                         & 164                                        & 117 (1.0\%)    & 133 (1.1\%)  & $< 0.01$ & $0.01$   \\
				SMT~\cite{smt-token}      & 2018-10376           & \num{1}                     & \num{56555}                         & 1                                  & 47                                         & 108                                        & 191 (0.8\%)    & 97  (0.4\%)  & $< 0.01$ & $0.01$   \\
				UET~\cite{uet-token}      & 2018-10468           & \num{55}                    & \num{24034}                         & 12                                 & 225                                        & 21                                         & 1,299 (18.2\%) & 541 (7.6\%)  & $< 0.01$ & $0.071$    \\
				SCA~\cite{sca-token}      & 2018-10706           & \num{1}                     & \num{292}                           & 10                                 & 47                                         & 0                                          & 3,811 (17.3\%) & 361 (1.6\%)  & $< 0.01$ & $0.189$    \\
				HXG~\cite{hxg-token}      & 2018-11239           & \num{9}                     & \num{1497}                          & 5                                  & 120                                        & 541                                        & 997 (28.1\%)   & 519 (14.6\%) & $< 0.01$ & $0.057$    \\
				\bottomrule
			\end{tabular}
		}
	\end{center}
	\caption{ERC-20 Token contracts investigated in depth with their respective CVE number, the number of patches introduced by \toolname, and the number of transactions replayed by \toolname's patch tester and the number of attack transactions identified while testing the patches.
We also give the average amount of overhead in gas consumption over all replayed transactions and overhead of contract size of the manual patched contracts (SM) and rewriter-generated patches (RW) and the overhead of the rewriter converted to US\$ (with a gas price of \SI{1}{\giga\wei} and \SI{235}{US\$/\ether};
For readability we only show the exact US\$ figures only if they are more than one cent).}%
	\label{tab:overhead}
\end{table*}

From those 14107 contracts, around 8000 involve transactions on the Ethereum network.
To generate a large and representative evaluation data set, we extracted all transactions sent to these contracts up to block \num{7755100} (May 13 2019) from the Ethereum blockchain resulting in \num{26385532} transactions.

Replaying those transactions with our patch tester shows that for \SI{95.5}{\percent} of all vulnerable contracts, \toolname's generated patch was compliant to all of the prior transactions associated with those contracts. 
For the remaining \SI{4.5}{\percent} of the investigated contracts, our patch rejected transactions for one of the following reasons:
\begin{inparaenum}[(1)]
	\item we successfully stopped a malicious transaction,
	\item the reported vulnerability was a false positive and should not have been patched, or
	\item we unintentionally changed the contract's functionality.
\end{inparaenum}

For close scrutiny, we selected ERC-20 token contracts from those contracts that could be successfully patched by \toolname\ with confirmed integer overflow/underflow vulnerabilities that have been successfully attacked (see \Cref{tab:overhead}).
For comparison purposes, we also manually patch these contracts on the Solidity source code level by replacing the vulnerable arithmetic operations with functions adapted from the \emph{SafeMath} library~\cite{safemath}.
The manually patched source code is then compiled with the exact same Solidity compiler version and optimization options used in the original contract (as reported on \href{https://etherscan.io/}{etherscan.io}).

We applied the \toolname\ patch tester to the generated patched contract versions and validated the reported outcome. %
This allows us to verify whether both patching approaches abort the same attack transactions.
In addition, we can compare the overhead in gas consumption and the increase in code size.
Note that in the manual patching method, we do not patch all potential vulnerabilities detected by \osiris as we skip adding checks on those arithmetic operations which cannot be exploited by an attacker, i.e., vulnerable arithmetic operations contained in functions that can only be called by the controller or owner of the contract. We verified the correctness of our patches using a total number of \num{506607} real-world transactions associated with the ERC-20 token contracts listed in \Cref{tab:overhead}.

Table~\ref{tab:overhead} shows the transaction execution results of the patch tester.
We verified the aborted transactions and confirm that all of them correspond to genuine attacks except for one transaction\footnote{\scriptsize 0x776da02ce8ce3cc882eb7f8104c31414f9fc756405745690bcf8df21e779e8a4}, which resembles a special case of token burning that we discuss in detail below.
Apart from the valid attack transactions, the execution traces of the re-executed transactions match those of the original transactions, confirming that our patch does not break the contract's functionality.

Out of the transactions identified as attacks, we found one particular transaction to the HXG token~\cite{hxg-token}.
The transaction does indeed trigger an integer overflow but the HXG token rather burns some tokens by transferring them to a blackhole address \hex{0x0}.
The burned tokens cannot be recovered and the balance of the blackhole address does not influence the behavior of the contract.
When analyzing the contract, \osiris\ is not aware of the semantics of this blackhole address and reports a possible integer overflow.
\toolname\ then conservatively patches the integer overflow bugs reported by \osiris, which leads to one legitimate transaction failing.
We argue that this pattern can be seen as bad coding practice as it wastes gas in unnecessarily storing the balance of the blackhole address.

\paragraph{Gas Overhead.}
The additional code introduced by the patching may potentially cause transactions to fail with an out-of-gas error.
While the patches generally do not significantly increase gas consumption, such a behavior can nevertheless occur when the sender of the transactions provides a very tight gas budget.
When the re-execution of a transaction with patched code fails early due to an out-of-gas exception, we could not accurately compare the behavior of the patched contract with the original contract.
To remedy this, we disabled the gas-accounting in the \evm.
We report the amount of additional gas consumption during transaction execution in Table~\ref{tab:overhead}.
We excluded those transactions that do not execute functions which contain the vulnerable code, because they are not affected by the patches and therefore not relevant to our measurements.

Our results show that for contracts BEC, SMT, and HXG, those patched with \toolname\ incur less gas overhead at runtime (\SIlist{83;47;120}{\gas}) when compared to those patched on the source code level (\SIlist{164;108;541}{\gas}).
This is due to the fact that the Solidity compiler generates non-optimal code when only very few checks are added.
In particular, Solidity utilizes internal function calls to invoke the SafeMath integer overflow checks. While this reduces code size (in case the check is needed at multiple places), it always requires executing additional instructions---thereby increasing gas overhead---to invoke and return from the internal function.
In contrast, \toolname\ inlines the safe numeric operations thereby introducing less gas overhead.
One would need to instruct the Solidity compiler to selectively enable function inlining to yield similar gas costs as \toolname.

Note that the average gas overhead is \SI{0}{\gas} for the manually patched SCA token. This is because only one transaction triggers the SafeMath integer overflow check. However, this is an attack transaction and it is aborted early, making gas overhead calculation not possible.

For UET and SCA, we identify higher gas overhead than for the manually patched version.
In fact, UET requires on average \num{255} units of additional gas for every transaction in the patched version. In contrast, only \SI{21}{\gas} is added for manually patched version.
This is due to the fact that our bytecode rewriter conservatively patches every potential vulnerability reported by \osiris\ in these two contracts (\num{12} and \num{10} respectively).
However, not all of them are actually exploitable and as such we did not instrument them during manual patching.

\paragraph{Code Size Increase.}
Deploying contracts in the Ethereum blockchain also incurs costs proportionally to the size of the deployed contract.
More specifically, Ethereum charges \SI{200}{\gas} per byte to store the contract code on the blockchain~\cite{Wood2016-yellowpaper}.
From Table~\ref{tab:overhead}, we recognize that the amount of extra code added by our rewriter is comparable to that of the SafeMath approach when a single vulnerability is patched.
Since our approach duplicates the original basic blocks, the code size overhead depends on the specific location of the vulnerability.
In the case of the BEC token contract, our rewriter increases the code size less than the source-level patches.
The Solidity compiler generates more code for including the SafeMath library than is strictly necessary for the patch.
Even considering the overhead of bytecode rewriting, we observe that \toolname\ generates a smaller patch than the manual patching method for this contract.

However, in case many vulnerabilities are patched, \toolname\ adds a slightly higher overhead.
Naturally, the size of the upgraded contracts increases with the number of vulnerabilities to fix due to inlining.
For instance, our bytecode rewriter generates 12 patches for UET contract and 10 patches for SCA contract resulting in \SI{1299}{\byte} (18.2\%) and \SI{3811}{\byte} (17.3\%) increase in code size.
In the worst-case scenario in our dataset, this increase in code size induces negligible additional cost of \usd{0.1791863602} per deployment.

Our patch templates are currently optimized for patching a single vulnerable arithmetic.
It is straightforward to adopt an approach akin to Solidity's internal function calls when developing patch templates for our bytecode rewriter, which would reduce the code size overhead when patching many integer overflows.
\toolname\ applies \(3.9\) patches on average to a contract in our data set of \num{14107} contracts.
The average code size of the original contracts is \SI{8142.7}{\byte} (\(\sigma\) \SI{5327.8}{\byte}).
The average size increase after applying patches with \toolname\ is \SI{455.9}{\byte} (\(\sigma\) \SI{333.5}{\byte}).
This amounts to an average code size overhead of 5.6\% after applying the patches.
Given that Ethereum charges \SI{200}{\gas} per byte to the contract creation transaction, it incurs an average overhead of \SI{91180}{\gas} or \usd{0.021} at the time of writing. %
In the worst case that we observed, \toolname\ incurs an overhead of \SI{199800}{\gas} at deployment, which at the time of writing only amounts to about \usd{0.04} additional deployment cost.
This shows that the overhead of applying patches with bytecode rewriting is negligible for contract deployment, especially when compared to the number of Ether possibly at stake.

\paragraph{Costs of Deployment.}
The deployment cost of a newly patched contract dominates the costs of operating a smart contract with \toolname.
However, additionally there is a transaction needed to switch the address of the logic contract.
Since the proxy pattern requires no state migration, this transaction requires a constant amount of gas.
The proxy contract we utilize in \toolname\ consumes \SI{43,167}{\gas} during a switchover transaction, i.e., about \usd{0.01}.
Currently, state migration is the most viable contract upgrade strategy besides the proxy pattern.
Prior work estimated that even with only \num{5000} ERC-20 holders, i.e., smart contract users, state migration will likely cost more than \usd{100} in the best case~\cite{trailofbits-migration}.
Hence, compared to the cost of migrating all data to a new contract, the \toolname's additional cost of \usd{0.01} is negligible.

\begin{figure}[t]
	\centering
	\includegraphics[width=0.9\linewidth]{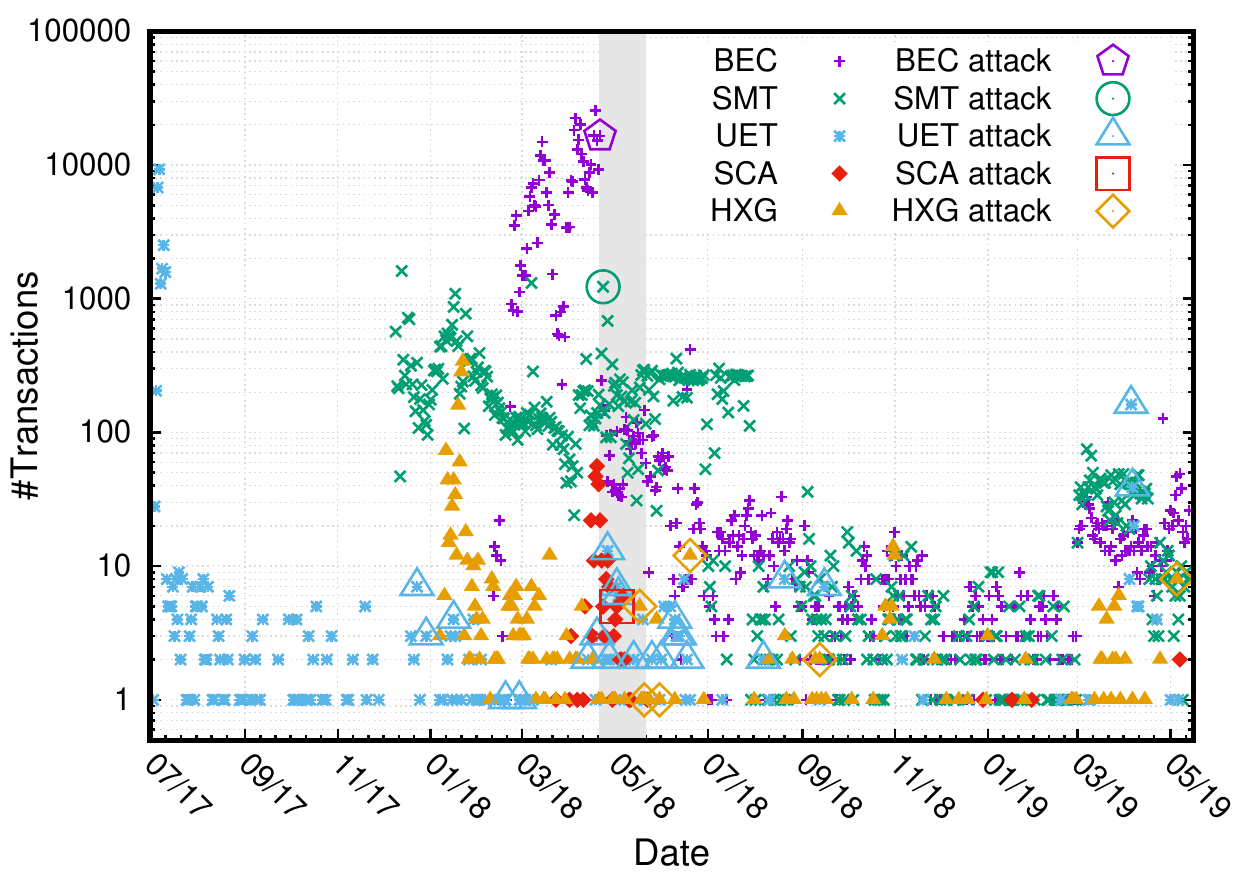}
	\caption{Activity timeline of each contract. The grey shadow indicates the
		time window in which the vulnerabilities of these contracts are disclosed
		by Peckshield~\cite{peckshieldadvisories}, and the big hollow points signify the occurrences of the attacks.}
	\label{fig:timeline}
\end{figure}

\paragraph{Detecting Attacks.}
The patch tester of \toolname\ allows us to also identify any prior attack transactions.
In Figure~\ref{fig:timeline}, we additionally observe that while the vulnerabilities of the other token contracts have been reported within a fairly reasonable time after the first attack, UET has been exploited (5 months) long before the bug disclosure.
More surprisingly, all contracts are still fairly active though they encountered a decrease of transaction volume after public disclosure of the vulnerabilities.
Despite the fact that all of these vulnerabilities have been discovered around one year before the time of writing, there are still \num{23630} transactions (\SI{4.66}{\percent} of the evaluated transactions) issued
to these vulnerable contracts after the public disclosure of the vulnerabilities, including successful
attacks.
This means that the owners of those contracts did not properly migrate to patched versions and users were not properly notified of the vulnerable state of these contracts.

\subsubsection{Analysis of False Positives/Negatives}%
\label{sec:evalfpfn}

During our analysis of the vulnerable contracts, we identified false positives and false negatives caused by vulnerability reporting of \osiris~\cite{osiris}.
This demonstrates that our patch testing is an important step in the process as many analysis tools are imprecise.
We found that in the default configuration, \osiris\ often achieves limited code coverage.
To this end, we utilized different timeout settings for both the whole analysis and for queries to the SMT solver and combined the results of multiple runs to achieve better code coverage.
Furthermore, we found that---contrary to the claims in the original \emph{\osiris} paper~\cite{osiris}---not all vulnerabilities are accurately detected by \osiris\ in two particular cases.

\paragraph{Hexagon (HXG) Token.}
This contract is vulnerable to an integer overflow, which allows an attacker to transfer very large amounts of ERC-20 tokens~\cite{burnOverflowHexagon}.
\osiris\ reports two false positives, which are caused by \evm\ code that is generated by the Solidity compiler.
Even though all types are unsigned types in the Solidity source code, the compiler generates a signed addition.
Here, \osiris\ reports a possible integer overflow, when \(-2\) is added to the \emph{balanceOf} mapping variable.
When performing signed integer additions with negative values, the addition naturally overflows when the result moves from the negative value range into the positive value range and vice versa.
As such, \toolname\ patches a checked addition for an unsigned arithmetic operation which will always overflow.
With our patch tester we observe all the failing transactions and perform manual analysis of the patched contract's bytecode to determine that the root cause is an issue in the Solidity compiler, i.e., the generated code requires an additional instruction, when compared to a simple unsigned subtraction.

\paragraph{Social Chain (SCA).}
Our results also show a problem with \osiris when analyzing the SCA token.
While \osiris\ does detect a possible overflow during multiplication in the problematic Solidity source code line, it does not detect the possible integer overflow for an addition in the same source code line.
However, in the actual attack transaction, the integer overflow happens during the not-flagged addition operation.
As such, this constitutes a false negative problem of \osiris.
Since the vulnerable addition is not reported by \osiris, it is also not automatically patched by \toolname.
In contrast, for the manually patched version we took both arithmetic operations into account.
The related attack transaction was previously reported as an attack transaction~\cite{SCAmultioverflow}.

\paragraph{Summary of Evaluation.}
To summarize, our evaluation on integer overflow detection shows that \toolname\ can correctly apply patches to smart contracts preventing \emph{any} integer overflow attack.
Furthermore, \toolname\ incurs only a \emph{negligible} gas overhead during deployment and runtime; especially compared to the Ether at stake.
Our analysis shows that the analyzed vulnerable smart contracts are \emph{still in active use, even after being attacked} and the vulnerabilities being publicly disclosed.
This motivates the need for a timely patching framework such as \toolname.
Lastly, based on an extensive and detailed analysis of \num{26385532} transactions, we demonstrate that \toolname\ always preserves the contract's original functionality except for a few cases, where the vulnerability report (generated by the third-party tool \osiris) was not accurate or bad coding practices were used (blackhole address).

\subsection{Developer Study}%
\label{sec:evaldevstudy}

\begin{table}[b]
	\begin{center}
		\caption{Timing results for the tasks as reported by the developers given in minutes and their reported confidence in the correctness of their results.}%
		\label{tab:devstudtimings}
		{\small
      \begin{tabulary}{\linewidth}{L c c c c}
        \toprule
        Task                            & \multicolumn{3}{c}{Time (Minutes)} & Confidence                            \\
                                        & Median                             & Min        & Max       & Median (1-7) \\
        \midrule
        Manual Integer Patches             & \num{47.50}                        & \num{35}   & \num{78}  & \num{6}      \\
        Conversion    & \num{62.50}                        & \num{33}   & \num{110} & \num{2.5}    \\
        \midrule
        \toolname\ Conversion & \num{1.50}                         & \num{1}    & \num{3}   & -            \\
        Patch Template        & \num{4.00}                         & \num{2}    & \num{15}  & \num{7}      \\
        \bottomrule
      \end{tabulary}
		}
	\end{center}
\end{table}

\paragraph{Developer Background.}
To quantify the manual effort needed to patch smart contracts and evaluate the usefulness of \toolname\, we conducted a thorough study with 6~professional developers with varying prior experience in using blockchain technologies and developing smart contracts.
Our developers consider themselves familiar with blockchain technologies but not very familiar with developing Solidity code.
None of the developers have developed an upgradable contract before.
As such, we can quantify the effort needed for a smart contract developer to learn and apply an upgradable contract pattern.

\paragraph{Methodology.}
Throughout our study, we asked the developers to perform multiple tasks manually that are performed automatically by \toolname:
\begin{inparaenum}[(1)]
	\item manually patch three contracts vulnerable due to integer overflow bugs given the output of a static analyzer (OSIRIS~\cite{osiris}),
	\item convert a contract to an upgradable contract manually and with \toolname, and
	\item patch an access control bug using \toolname\ by writing a custom patch-template.
\end{inparaenum}
The three tasks cover different scenarios, where \toolname\ can be useful to a developer.
The first two tasks cover the use of \toolname\ to patch known bug classes with minimal human intervention.
For these two tasks we assume no prior knowledge on patching smart contracts (see \Cref{tab:devstudyquest} how developers rated their prior experience with smart contracts).
In contrast, the third task consists of extending \toolname. 
This requires understanding a bug class and perform root cause analysis to properly patch the vulnerability.
This is surely more challenging compared to the previous two tasks.
Since the third task covers a different bug class, we believe there is no significant bias in the data due to the developers completing the other two tasks first.

For all tasks, we measured the time required by the developer to perform the task (excluding the time required for reading the tasks' description).
We asked the developers to rate their familiarity with relevant technologies, their confidence levels in their patches, and the difficulty of performing the tasks on a 7-point Likert scale.
The full questionnaire and the answers of the developers are shown in \Cref{tab:devstudyquest}, and the recorded time measurements are shown in \Cref{tab:devstudtimings}.
We provide the supporting files in a github repository.\footnote{\href{https://github.com/uni-due-syssec/evmpatch-developer-study}{github.com/uni-due-syssec/evmpatch-developer-study}}

We then performed both a manual code review and a cross-check with \toolname\ to analyze mistakes made by the developers.
The results of our study show that significant effort is needed to correctly patch smart contracts manually, whereas \toolname\ enables simple, user-friendly, and efficient patching.
The time measurements show that the developers, who had no prior experience with \toolname, were able to perform complex tasks utilizing \toolname\ within minutes.

\paragraph{Patching Integer Overflow Bugs.}
We asked the developers to fix all integer overflow vulnerabilities in three contracts:
\begin{inparaenum}[1]
	\item BEC~\cite{bec-token} (CVE-2018-10299, 299 lines of code), and
	\item HXG~\cite{hxg-token} (CVE-2018-11239, 102 lines of code) and
	\item SCA~\cite{sca-token} (CVE-2018-10706, 404 lines of code).
\end{inparaenum}
To provide a representative set of contracts, we chose three ERC-20 contracts with varying complexity (in terms of lines of code) and where the static analysis also includes missed bugs and false alarms (see \Cref{sec:evalfpfn}).
We ran OSIRIS on all three contracts and provided the developers the analysis output as well as a copy of the SafeMath Solidity library.
This accurately resembles a real-world scenario, where a blockchain developer quickly needs to patch a smart contract based on the analysis results of recent state-of-the-art vulnerability analysis tools and can look-up manual patching tutorials available online.
All developers manually and correctly patched the source code of all three contracts which demonstrates their expertise in blockchain development.
However, on the downside, it took the developers on average \SI{51.8}{\minute} ($\sigma = \SI{16.6}{\minute}$) to create patched version for the three contracts.
In contrast, \toolname\ fully automates the patching process and is able to generate patches for the three contracts within a maximum of \SI{10}{\second}.

\paragraph{Converting to an Upgradable Contract.}
The developers had to convert a given smart contract into an upgradable smart contract.
We provided the developers a short description of the delegatecall-proxy pattern and asked them to convert the given contract into two contracts: one proxy contract and a logic contract, which is based on the original contract.
We provided no further information on how to handle the storage-layout problem, and we explicitly allowed using code found online.
The developers required an average of \SI{66.3}{\minute}\footnote{$\sigma = \SI{31.3}{\minute}$, fastest \SI{33}{\minute} and slowest \SI{110}{\minute}} to convert a contract into an upgradable contract.
\emph{None} of the developers performed a correct conversion into an upgradable contract, which is also reflected in a median confidence of \num{2.5} in the correctness reported by the developers.
We observed two major mistakes:
\begin{inparaenum}[(a)]
	\item The proxy contract would only support a fixed set of functions, i.e., the proxy would not support adding functions to the contract, and
	\item more importantly, only one out of six developers correctly handled storage collisions in the proxy and logic contract, i.e., five of the six converted contracts were broken by design.
\end{inparaenum}
Hence, it remains open how long it would take developers to perform a correct conversion. %

Next, we asked the developers to utilize \toolname\ to create and deploy an upgradable contract.
As \toolname\ does not require \emph{any} prior knowledge about upgradable contracts, the developers were able to deploy a correct upgradable contract within at most \SI{3}{\minute}.
In addition, patching with \toolname\ inspires high confidence---a median of \num{7}, the best rating on our scale---in the correctness of the patch.
This gives a strong confirmation that deployment of a proxy with \toolname\ is indeed superior to manual patching and upgrading.

\paragraph{Extending \toolname.}
The developers had to write a custom patch template for \toolname.
We instructed the developers on how to use \toolname\ and how patch templates are written with \toolname's patch template language (see \Cref{fig:parity-patch} for an example).
Furthermore, we presented the developers an extended bug report that shows how an access control bug can be exploited.
The developers leveraged the full \toolname\ system, i.e., \toolname\ applies the patch and validates the patch using the patch tester component which replays past transactions from the blockchain and notifies the developer whether:
\begin{inparaenum}[(a)]
	\item the patch prevents a known attack, and
	\item whether the patch broke functionality in other prior legitimate transactions.
\end{inparaenum}
As such, \toolname\ allowed the developers to create a fully functional and securely patched upgradable contract within a few minutes.
On average, the developers only needed \SI{5.5}{\minute}, and a maximum of \SI{15}{\minute}, to create a custom patch template.
As expected, all developers correctly patched the given contract using \toolname, because a faulty patch would have been reported by \toolname's patch tester to the developer.
\toolname's integrated patch tester gives the developers a high confidence into their patch.
On average, the developers reported a confidence level of \num{6.6} ($\sigma = 0.4$), where 7 is the most confident.
Furthermore, \emph{none} of the developers considered writing such a custom patch template as particularly difficult.

\paragraph{Summary.}
Our study provides confirmation that \toolname\ offers a high degree of automation, efficiency, and usability thereby freeing developers from manual and error-prone tasks.
In particular, none of the six developers were able to produce a correct upgradable contract mainly due to the difficulty of preserving the storage-layout.
Our study also confirms that extending \toolname\ with custom patch templates is a feasible task, even for developers that are unaware of the inner workings of \toolname.

\section{Related Work}%
\label{sec:relatedwork}

The infamous attack against \theDAO\ contract~\cite{dao-attack-analysis} received considerable attention from the community.
Since then, many additional exploits and defenses, which mostly focus on discovering bugs before the contract is deployed, were revealed.
Luu et al.\ presented the symbolic executor Oyente that explores a contracts code, while looking for possible vulnerabilities~\cite{luu2016making}.
Since then many other symbolic execution tools with better precision, performance, and covering different vulnerabilities have been proposed~\cite{mythril,osiris,Krupp2018-teether,manticore-paper,maian}.
Furthermore, static analyzers for both Solidity~\cite{slither} and \evm\ bytecode have been proposed~\cite{Tsankov2018-lq}.
Information flow analysis and data sanitization in a multi-transaction setting is analyzed by Ethainter~\cite{Brent2020ethainter}.
Furthermore, methods from formal verification and model checking have been applied to smart contracts~\cite{zeus-ndss2018,Frank2020ethbmc} and the semantics of the \evm\ and Solidity language have been formalized~\cite{Grishchenko2018-pm,Jiao2020-js}.
However, only a small body of prior work has researched dynamic analysis and runtime protections.
Tools such as Sereum~\cite{sereum-ndss19} or ECFChecker~\cite{ecfchecker} can detect live reentrancy attacks on vulnerable contracts.
Recent work has further explored modular dynamic analysis frameworks for protecting smart contracts~\cite{Chen2020soda,Torres2020aegis}.
Protection solutions that require modifications to the smart contract execution environment are unlikely to be integrated in to production blockchain systems.

Integer overflows have been widely studied in the context of Ethereum smart contracts.
\osiris~\cite{osiris} is an extension to the symbolic execution tool Oyente~\cite{luu2016making} to accurately detect integer bugs.
The improved symbolic execution engine first attempts to infer the integer type, i.e., signedness and bit width, from the specific instructions generated by Solidity compilers.
Next, it checks for possible integer bugs, such as truncation, overflow, underflow, and wrong type casts.
We leverage the detection capabilities of \osiris, because it pinpoints the exact location of the integer overflow bug.
Other tools such as TeEther~\cite{Krupp2018-teether} and MAIAN~\cite{maian} implicitly find integer bugs when they generate exploits for smart contracts.
However, they do not report the exact location of the integer overflow, because they focus on exploit generation.
ZEUS~\cite{zeus-ndss2018} utilizes abstract interpretation and symbolic model checking to verify safety properties of smart contracts.
While ZEUS can detect potential integer overflow vulnerabilities, it does so at the LLVM intermediate level and cannot determine the exact location in the corresponding \evm\ bytecode.

Recently, bytecode rewriting for patching smart contracts has been explored with \emph{SMARTSHIELD}~\cite{Zhang2020smartshield}. 
SMARTSHIELD requires a complete control-flow graph (CFG) to update jump targets and data references.
As discussed in \Cref{sec:challenges}, generating a highly accurate CFG is highly challenging due to the \evm's bytecode format.
We believe that such a bytecode rewriting strategy does not scale to larger and more complicated contracts.
In contrast, \toolname's trampoline-based rewriting strategy does not require an accurate CFG and is much more resilient when rewriting complex contracts.
SMARTSHIELD implements custom bytecode analysis to detect vulnerabilities, which may not be as accurate as specialized analyses.
For example, SMARTSHIELD's analysis does not infer whether an integer type is signed, which is important for accurate integer overflow detection~\cite{osiris}.
\toolname\ is a flexible framework that can integrate many static analysis tools for detecting vulnerabilities and can leverage analysis tool improvements with minimal effort.
Last and most importantly, \toolname\ automates the whole lifecycle of deploying and managing an upgradable contract, while SMARTSHIELD is designed to harden a contract pre-deployment.
With \toolname, a smart contract developer can also patch vulnerabilities that are discovered after deployment of the contract.

The Ethereum community explored several design patterns to allow upgradable smart contracts~\cite{upgradablecontracts,trailofbits-migration,zeppelinos-docs,zeppelinos-blog-proxy-patterns}  with manual migration to a new contract and the proxy pattern being the most popular (see \Cref{sec:upgrade}).
The ZeppelinOS~\cite{zeppelinos-docs} framework supports upgradable contracts by implementing the delegatecall-proxy pattern.
However, developers have to manually ensure compatibility of the legacy and patched contract on the Solidity level.
This can be achieved using static analysis tools that perform ``upgradeability'' checks (e.g., Slither~\cite{slither-upgrade-checks} checks for a compatible storage layout), which relies on accurate knowledge of compiler behavior with respect to storage allocations.
On the other hand, \toolname\ combines existing analysis tools and provides an automatic method to patch detected vulnerabilities while keeping storage layout consistent by design.
\section{Conclusion}%
\label{sec:conclusion}

Updating erroneous smart contracts constitutes one of the major challenges in the field of blockchain technologies.
The recent past has shown that attackers are fast in successfully abusing smart contract errors due to the natural design of the underlying technology: always online and available, one common and simple computing engine without any subtle software and configuration dependencies, and (often) high amount of cryptocurrency at disposal.
While many proposals have introduced frameworks to aid developers in finding bugs, it remains open how developers and the community can quickly and automatically react to vulnerabilities on already deployed contracts.
In this work, we developed a framework that supports automated and instant patching of smart contract errors based on bytecode rewriting.
In terms of evaluation, we were able to demonstrate that real-world vulnerable contracts can be successfully patched without violating the functional correctness of the smart contract.
Our developer study shows that an automated patching approach greatly reduces the time required for patching smart contracts and that our implementation, \toolname, can be practically integrated into a smart contract developers workflow.
We believe that automated patching will increase the trustworthiness and acceptance of smart contracts as it allows developers to quickly react on reported vulnerabilities.

\section*{Acknowledgment}
The authors would like to thank the reviewers---and especially our shepherd Yinzhi Cao---for their valuable feedback, and the developers for taking the time to participate in our study.
This work was partially funded by the Deutsche Forschungsgemeinschaft (DFG, German Research Foundation) under Germany's Excellence Strategy - {EXC} 2092 {CASA} - 390781972 and the DFG as part of project S2 within the CRC 1119 CROSSING.
This work has been partially supported by the EU {H2020-SU-ICT-03-2018} CyberSec4Europe project, funded by the European Commission under grant agreement no. 830929.

{
  \footnotesize

  \printbibliography
}

\appendix%

\section{Rewriter Example}%
\label{sec:rewriterexample}

\begin{figure}[t!]
  \begin{center}
    \includegraphics[width=0.8\linewidth]{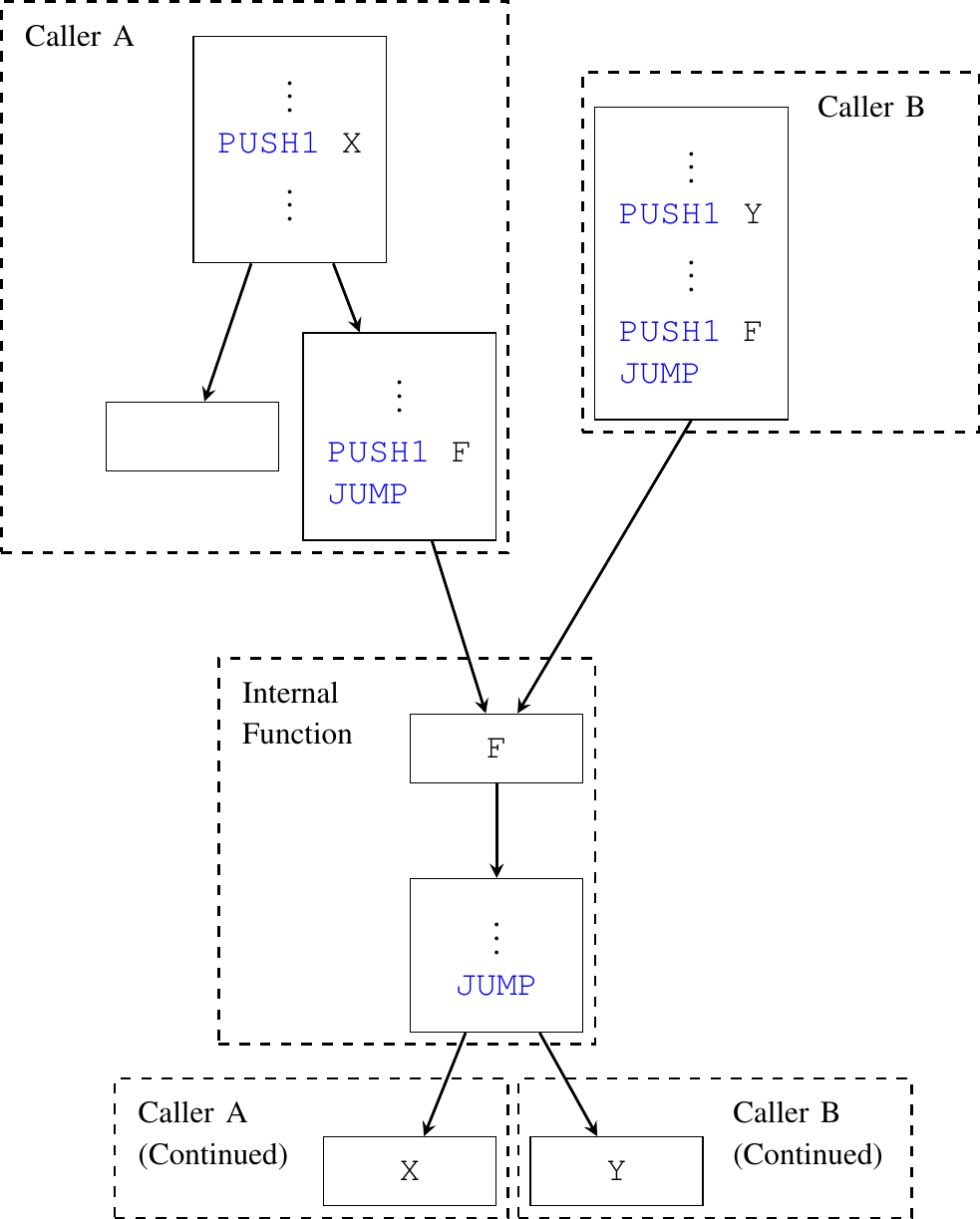}
  \end{center}%
  \caption{A jump when calling an internal function in Solidity. The actual jump target is a constant which is pushed several instructions before the jump is executed. Furthermore, the jump target is typically masked using the \texttt{AND} instruction. The basic block leading to the jump must be emulated to compute the actual jump target.}%
  \label{fig:jumpinternalcall}
\end{figure}

Consider Figure~\ref{fig:jumpinternalcall} which depicts an excerpt of a typical \cfg\ when Solidity generates code utilizing internal function calls.
There are two callers of the internal function, named \emph{A} and \emph{B}.
Each caller pushes a constant return address onto the stack: \emph{A} pushes the return address \emph{X}; \emph{B} the constant \emph{Y}.
These constants are addresses of basic blocks, where execution resumes once the internal function completes.
In this example, both callers push the address of the first basic block of the internal function, dubbed \emph{F}, onto the stack and utilize the jump instruction to call the internal function.

To emulate the return, an indirect jump instruction is leveraged at the end of the internal function, where the target address is taken from the stack.
Depending on the calling context, the final jump of the internal function will either jump back into \emph{A} or \emph{B}.
Note that, for this example, the surrounding basic block does not contain any corresponding push instruction of the jump target.
Instead, the respective push instruction issued at the call site has loaded the return address on the stack.
Hence, data-flow analysis is needed to determine push instructions that are leveraged for function returns.

\section{Checked Add Example}%
\label{sec:checkedaddexample}

\begin{figure}[t]
\begin{center}
  \includegraphics[width=0.4\textwidth]{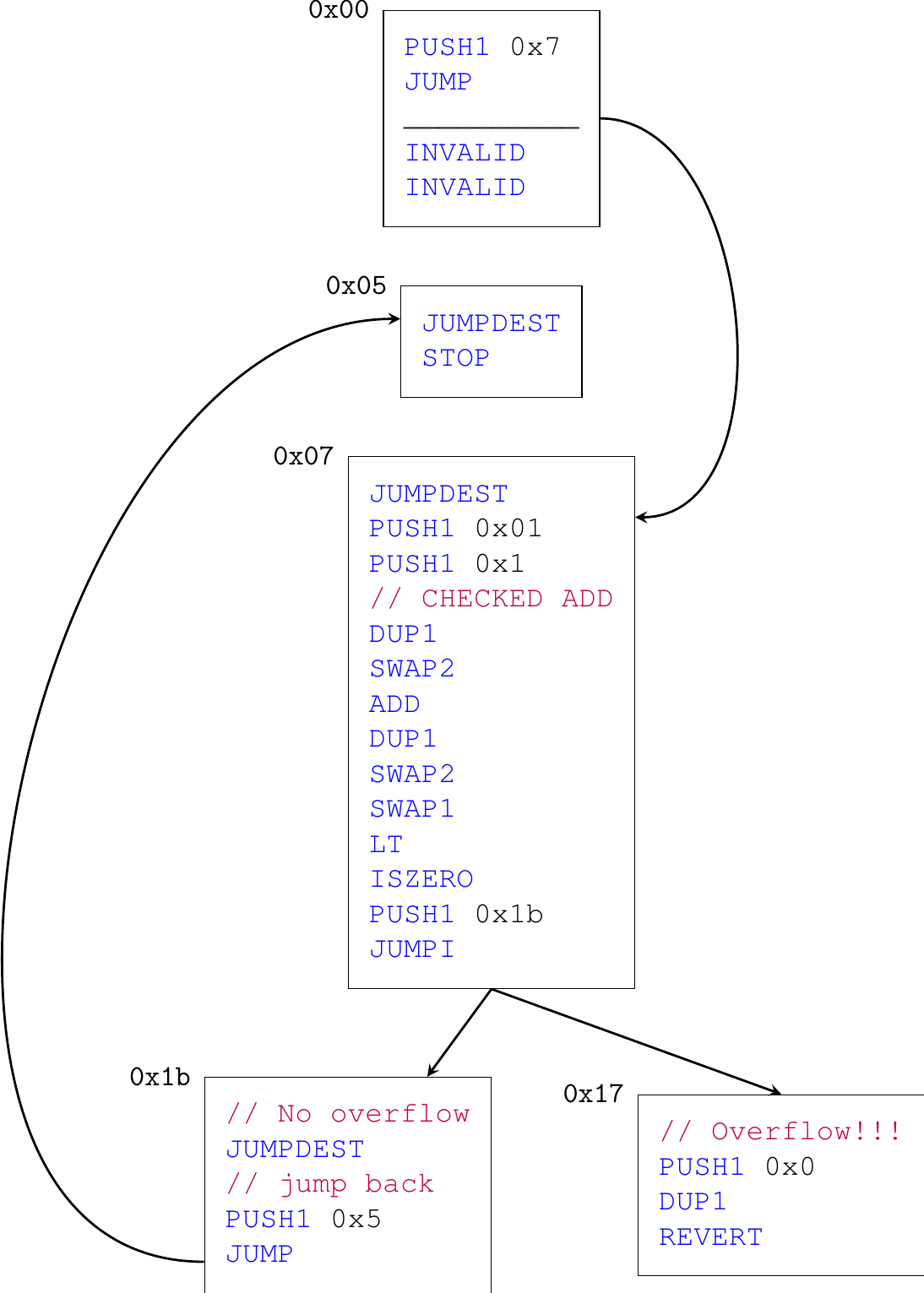}
\end{center}
\caption{Control-Flow Graph of the rewritten code with the checked addition routine included.}
\label{fig:checkedaddcfg}
\end{figure}

This sections discusses an example for an original and patched version of the same bytecode, as produced by our bytecode rewriter.
Figure~\ref{fig:addexample} shows the original target contract, which simply adds two constant numbers and then stops execution.
In this case the \eInst{ADD} instruction at address \hex{0x04} will be replaced with a checked addition routine.
Figure~\ref{fig:checkedaddexample} shows the linear disassembly of the patched code and Figure~\ref{fig:checkedaddcfg} shows the control-flow graph (CFG) of the rewritten code.
The original basic block at address \hex{0x00} has been replaced with a trampoline in the patched code.
The trampoline immediately jumps to the patched copy of the original basic block at address \hex{0x07}.
Here, the checked addition patch template has replaced the original \eInst{ADD} instruction.
Note that the rewriter not only modified the original basic block, but also introduced additional basic blocks and edges into the \cfg.
The inserted trampoline code and the jump-back to the original code amount to 23 additional gas used by the contract.
This is the overhead introduced by the bytecode rewriter.
In this example, we apply a checked add patch template, which verifies the invariant that \((a + b) >= a\).
The checked version of the addition requires an additional 35 gas when no overflow is detected.
This sums up to a total overhead of 58 gas.

\begin{figure}[t]
\begin{center}
\begin{lstlisting}[language=evm,xleftmargin=3em]
// original code
0x00: PUSH1 0x1  // (gas: 3)
0x02: PUSH1 0x1  // (gas: 3)
0x04: ADD        // (gas: 3)
0x05: JUMPDEST   // (gas: 1)
0x06: STOP       // (gas: 0)
\end{lstlisting}
\end{center}
\caption{Original code, which simply adds two numbers. The comments show the gas utilization of the instructions.}%
\label{fig:addexample}
\end{figure}

\begin{figure}[t]
\begin{center}
\begin{lstlisting}[language=evm,xleftmargin=3em]
// patched code
// TRAMPOLINE
0x00: PUSH1 0x7  // (gas: 3)
0x02: JUMP       // (gas: 8)
0x03: INVALID    // (gas: 0)
0x04: INVALID    // (gas: 0)
// continuation of original code
0x05: JUMPDEST   // (gas: 1)
0x06: STOP       // (gas: 0)
// -------------------------------------------------------
// PATCHED BASIC BLOCK (called via trampoline)
0x07: JUMPDEST   // (gas: 1)
0x08: PUSH1 0x1  // (gas: 3)
0x0a: PUSH1 0x1  // (gas: 3)
// CHECKED ADD
0x0c: DUP1       // (gas: 3)
0x0d: SWAP2      // (gas: 3)
0x0e: ADD        // (gas: 3)
0x0f: DUP1       // (gas: 3)
0x10: SWAP2      // (gas: 3)
0x11: SWAP1      // (gas: 3)
0x12: LT         // (gas: 3)
0x13: ISZERO     // (gas: 3)
0x14: PUSH1 0x1b // (gas: 3)
0x16: JUMPI      // (gas: 10)
0x17: PUSH1 0x0  // (gas: 3)
0x19: DUP1       // (gas: 3)
0x1a: REVERT     // (gas: 0)
0x1b: JUMPDEST   // (gas: 1)
// jump-back to original code:
0x1c: PUSH1 0x5  // (gas: 3)
0x1e: JUMP       // (gas: 8)
\end{lstlisting}
\end{center}
  \caption{Rewritten code, the first basic block is replaced with a trampoline jumping to the patched copy of the basic block at the end. The patched copied now performs a checked addition. The comments show the gas utilization of the instructions.}%
\label{fig:checkedaddexample}
\end{figure}

\section{Detailed Analysis of False-Positives of \osiris\ in Hexagon Token}%
\label{sec:hexagondetails}

The Hexagon Token contract is vulnerable to an integer overflow, which allows an attacker to transfer very large amounts of ERC-20 Tokens~\cite{burnOverflowHexagon}.
\osiris\ reports two false positives, which are caused by \evm\ code that is generated by the Solidity compiler.
Figure~\ref{fig:burnoverflowlisting} shows the Solidity and corresponding \evm\ code.
In the Solidity code in the upper listing of Figure~\ref{fig:burnoverflowlisting}, we can see that the variable \verb+_value+ is of type unsigned (Line 6) and the variable \texttt{burnPerTransaction} is also unsigned (Line 1).
Even though all types are unsigned types in the addition in Line 9, the compiler generates a signed addition.
The signed addition can be seen in lines 7 to 9 in the lower listing, where a negative signed value is pushed onto the stack.
Here, \osiris\ reports a possible integer overflow, when \(-2\) is added to the \emph{balanceOf} mapping variable.
When performing signed integer additions with negative values, the addition naturally overflows when the result moves from the negative value range into the positive value range and vice versa.

When patching the \eInst{ADD} on line 12 in the lower listing of Figure~\ref{fig:burnoverflowlisting}, with a checked addition, we introduce a false positive.
Replacing a signed addition with a checked unsigned addition will always fail if negative numbers are involved, since they naturally trigger a overflow when switching between the positive and negative ranges due to the two's complement representation.
The patch tester in our pipeline marked almost all transactions as failing, which is a strong indicator for a failed patch.

\begin{figure}[ht]
\begin{center}
  \input{figures/burnoverflowhexagon_fp.tex}
\end{center}
\caption{Problematic Solidity line in the Hexagon contract (top listing). Solidity generates the \evm\ code in the bottom listing. Instead of subtracting \(2\) from an unsigned integer, Solidity promotes this to an signed integer and adds \(-2\).}%
\label{fig:burnoverflowlisting}
\end{figure}

\section{Developer Study Questionnaire}

\Cref{tab:devstudyquest} shows the full questionnaire and the corresponding answers of the developers.
The description of the tasks and the document we provided alongside the questionnaire can be found on our github page: \href{https://github.com/uni-due-syssec/evmpatch-developer-study}{github.com/uni-due-syssec/evmpatch-developer-study}

\begin{table*}[b]
	\begin{center}
		\caption{Developer study questionnaire and answers by six developers (identified by the letters A to F).}%
		\label{tab:devstudyquest}
		{\footnotesize
\begin{tabulary}{\linewidth}{>{\bfseries}l L c c c c c c >{\bfseries}c R}
	\toprule
	\multicolumn{2}{c}{\textbf{Question}} & \multicolumn{7}{c}{\textbf{Answers}} & \textbf{Scale} \\
	\midrule
	& & A & B & C & D & E & F & Median & \\
	\midrule
  Q1 & Did you write Solidity code in the last two weeks? &	no &	no &	no &	no &	yes &	no & & (yes/no)\\
	Q2 & Have you previously worked on a production-grade Solidity-based Ethereum contract? & yes & no & no & no & no & no & & (yes/no)\\
	Q3 & Have you previously worked on a production-grade smart contract on another Blockchain Platform? & no & no & yes & no & yes & yes & & (yes/no)\\
	Q4 & How familiar are you with Blockchain technologies in general? & 6 & 5 & 7 & 6 & 6 & 6 & 6 & (1 not familiar, 7 very familiar)\\
  Q5 & How familiar are you with the Ethereum Blockchain in particular? & 6 & 5 & 4 & 2 & 6 & 2 & {4.5} &(1 not familiar, 7 very familiar)\\
	Q6 & How familiar are you with the Solidity programming language? & 6 & 3 & 2 & 1 & 5 & 1 & {2.5} & (1 not familiar, 7 very familiar)\\
	Q7 & How familiar are you with upgradable contracts in Solidity? & 5 & 3 & 1 & 1 & 4 & 1 & {2} & (1 not familiar, 7 very familiar)\\
	\midrule
	\multicolumn{10}{c}{Task 1} \\
	\midrule
	T1Q1 & How confident are you in the correctness of your patch to contract 1? & 5 & 7 & 7 & 6 & 7 & 6 & {6.5} & (1 least confident, 7 most confident)\\
	T1Q2 & How confident are you in the correctness of your patch to contract 2? & 6 & 7 & 7 & 4 & 7 & 6 & {6.5} & (1 least confident, 7 most confident)\\
	T1Q3 & How confident are you in the correctness of your patch to contract 3? & 3 & 5 & 6 & 5 & 2 & 4 & {4.5} & (1 least confident, 7 most confident)\\
	T1Q4 & How much time did you need to patch all three contracts? & 78 & 35 & 40 & 40 & 55 & 63 & {47.5} & (Time in Minutes) \\
	\midrule
	\multicolumn{10}{c}{Task 2} \\
	\midrule
  T2Q1 & Have you previously used the delegatecall-proxy pattern in a Solidity contract? & no & no & no & no & no & no& &(yes/no)\\
  T2Q2 & Have you previously used a different pattern to make a Solidity contract upgradable? & no & no & no & no & no & no& &(yes/no)\\
  T2Q3 & Have you previously used a different upgradable smart contract? & no & no & no & no & no & no& & (yes/no)\\
	T2Q4 & How confident are you in the correctness of your conversion? & 5 & 3 & 1 & 1 & 5 & 2 & {2.5} & (1 least confident, 7 most confident) \\
	T2Q5 & How difficult was the manual conversion? & 4 & 5 & 5 & 6 & 4 & 6 & {5} & (1 easy, 7 most difficult)\\
  T2Q6 & How difficult was the conversion using the evmpatch tool?  & 1 & 1 & 1 & 1 & 1 & 1 & {1} & (1 easy, 7 most difficult) \\
	T2Q8 & How much time did you need to convert the contract to an upgradable contract (Step 1)? & 110 & 80 & 45 & 90 & 40 & 33  & {62.5} & (Time in Minutes) \\
	T2Q8 & How much time did you need to convert the contract using EVMPatch (Step 2)? & 3 & 1 & 1 & 2 & 3 & 1 & {1.5} & (Time in Minutes) \\
	\midrule
	\multicolumn{10}{c}{Task 3} \\
	\midrule
	T3Q1 & How confident are you in the correctness of your patch? & 6 & 7 & 7 & 7 & 7 & 6 & {7} & (1 least confident, 7 most confident)\\
	T3Q2 & How difficult was the conversion using the EVMPatch tool?  & 2 & 1 & 1 & 1 & 1 & 1 & {1} & (1 easy, 7 most difficult) \\
	T3Q3 & How much time did you need to create and deploy the patch using EVMPatch? & 15 & 2 & 5 & 2 & 6 & 3 & {4} & (Time in Minutes)\\
	\bottomrule
\end{tabulary}
}
	\end{center}
\end{table*}

\end{document}